\newif\ifarxiv
\arxivtrue

\newif\ifdraft
\draftfalse

\documentclass[a4paper]{svjour3}

\usepackage[numbers]{natbib}

\usepackage{times}

\usepackage[includeheadfoot, head=13pt, foot=2pc, top=58pt, bottom=44pt, inner=60pt, outer=60pt, marginparwidth=2pc, heightrounded]{geometry}

\def\makeheadbox{}

\usepackage{mathpazo}
\usepackage[T1]{fontenc}
\usepackage[utf8]{inputenc}
\usepackage[scaled=0.81]{beramono}  

\usepackage{fontawesome}

\usepackage[multiple]{footmisc} \usepackage{tabularx,booktabs,multirow}
\usepackage[yyyymmdd]{datetime}

\usepackage{listings}
\usepackage{tabularx}

\usepackage{graphicx}
\graphicspath{{.}{figs/}}
\DeclareGraphicsExtensions{.png,.pdf}

\usepackage{tikz}
\usetikzlibrary{positioning,shapes.geometric,arrows,calc}
\tikzset{var/.style={draw=none,circle,fill=none,inner sep=3pt}}
\tikzset{plain/.style={var,fill=none}}
\tikzset{outcome/.style={var,draw=palcola,fill=none,ultra thick,text=black}}
\tikzset{treatment/.style={plain,fill=palcola!70,text=white}}
\tikzset{confounder/.style={plain,fill=palcolc,text=white}}
\tikzset{unmeasured/.style={plain,fill=palcolb,text=white}}

\usepackage{pgfkeys}

\usepackage{xparse,xstring,xspace}
\usepackage{dcolumn}
\usepackage{caption,subcaption}

\DeclareRobustCommand{\var}[1]{\ensuremath{\textnormal{\textsl{#1}}}}

\DeclareRobustCommand{\dist}[1]{\ensuremath{\textsf{#1}}}

\DeclareRobustCommand{\smd}[2]{\ensuremath{\textnormal{\textsc{scd}}(\var{#1} \to \var{#2})}}
\DeclareRobustCommand{\SMD}{\textsc{scd}\xspace}

\DeclareRobustCommand{\ev}[1]{\ensuremath{\mathds{E}[#1]}}
\DeclareRobustCommand{\sd}[1]{\ensuremath{\sigma[#1]}}

\usepackage{xcolor}
\definecolor{palcola}{HTML}{1b9e77}
\definecolor{palcolb}{HTML}{d95f02}
\definecolor{palcolc}{HTML}{7570b3}
\definecolor{palcold}{HTML}{e7298a}
\definecolor{palcole}{HTML}{66a61e}
\definecolor{palcolf}{HTML}{e6ab02}

\definecolor{degascola}{HTML}{591d06}
\definecolor{degascolb}{HTML}{96410e}
\definecolor{degascolc}{HTML}{e5a335}
\definecolor{degascold}{HTML}{556219}
\definecolor{degascole}{HTML}{418979}
\definecolor{degascolf}{HTML}{2b614e}
\definecolor{degascolg}{HTML}{053c29}

\newcommand{\suppOK}[1][green]{\text{\color{#1}\faCheck}}
\newcommand{\suppNO}[1][red]{\text{\color{#1}\faClose}}

\DeclareDocumentCommand{\gtmark}{O{degascolc}}{\textcolor{#1}{\faClose}\xspace}
\DeclareDocumentCommand{\estmark}{O{degascole}}{\textcolor{#1}{\textbullet}\xspace}

\DeclareDocumentCommand{\errorbar}{O{black} O{0.5ex} O{3pt}}{\tikz[baseline,yshift=#3] {
    \draw[#1, line width=1.5pt] (0, #2) -- (0, -#2);
    \draw[#1, line width=1pt] (-0.4ex, #2) -- (0.4ex, #2);
    \draw[#1, line width=1pt] (-0.4ex, -#2) -- (0.4ex, -#2);
  }\xspace }

\ifdraft
\usepackage{lineno}
\fi

\usepackage{algorithm2e}

\usepackage[inline]{enumitem}
\setlist[enumerate]{label=\emph{\roman*})}

\usepackage{amsmath}
\usepackage{graphicx}

\usepackage{xspace}

\usepackage{etoolbox}

\usepackage[colorinlistoftodos,textsize=tiny,obeyFinal,textwidth=16mm]{todonotes}
\NewDocumentCommand{\ColorComment}{O{color=white} m}{\todo[size=\scriptsize,#1]{#2}}

\NewDocumentCommand{\caf}{s m}{\ColorComment[\IfBooleanT{#1}{inline},color=purple!60]{#2}}
\NewDocumentCommand{\rfe}{s m}{\ColorComment[\IfBooleanT{#1}{inline},color=yellow]{#2}}
\NewDocumentCommand{\rto}{s m}{\ColorComment[\IfBooleanT{#1}{inline},color=green!70!black]{#2}}
\NewDocumentCommand{\msh}{s m}{\ColorComment[\IfBooleanT{#1}{inline},color=orange!70!black]{#2}}

\title{Mitigating Omitted Variable Bias in Empirical Software Engineering}

\author{
  Carlo A. Furia \and Richard Torkar
}
\institute{
  Carlo A.\ Furia \at
  Software Institute, USI Università della Svizzera italiana, Lugano, Switzerland\\
  \email{bugcounting.net}
  \and
  Richard Torkar \at
  Chalmers University of Technology and University of Gothenburg, Gothenburg, Sweden\\
  Stellenbosch Institute for Advanced Study (STIAS), Stellenbosch, South Africa\\
  \email{www.torkar.se}
}

\date{}

\makeatletter
\let\Title\@title
\makeatother

\usepackage{silence}
\usepackage{amsmath,amsfonts,dsfont}
\usepackage{mathtools}

\usepackage[colorlinks=true, allcolors=blue]{hyperref}

\hypersetup{
pdftitle={\Title{}},
pdfauthor={Carlo A. Furia, Richard Torkar},
pdfsubject={},
pdfkeywords={},
pdfcreator={pdfTeX}
}

\begin{document}

\maketitle

\ifdraft
  \linenumbers
\fi

\begin{abstract}
  Omitted variable bias occurs when a statistical model leaves out
  variables that are relevant determinants of the studied effects.
  This results in the model attributing the missing variables' effect
  to some of the included variables---hence over- or under-estimating 
  the latter's true effect.  Omitted variable bias presents a
  significant threat to the validity of empirical research,
  particularly in non-experimental studies such as those common in
  empirical software engineering.

  This paper illustrates the impact of omitted variable bias
  on two illustrative examples in the software engineering domain,
  and uses them to present methods
  to investigate the possible presence of omitted variable bias,
  to estimate its impact,
  and to mitigate its drawbacks.
  The analysis techniques we present are based on
  causal structural models of the variables of interest,
  which provide a practical, intuitive summary of the key relations among variables.

  This paper demonstrates 
  a sequence of analysis steps that inform
  the design and execution of similar empirical studies in software engineering.
  An important observation is that it pays off
  to invest effort investigating omitted variable bias
  \emph{before} actually executing an empirical study,
  because this effort can lead to a more solid study design,
  and to a reduction in its threats to validity.

\end{abstract}

\section{Introduction}
\label{sec:intro}

In this day and era of big data analytics,
where massive datasets are commonly available
with scalable machine learning techniques to boot,
it is tempting to believe that ``more data''
is a cure-all.
Unfortunately, there are scenarios where
training with more data cannot improve the quality
of a statistical analysis---in fact, it may even \emph{worsen} it.

Consider a common task of any empirical discipline:
estimating the (average) \emph{effect} that
changing a variable $X$ has on another, dependent variable $Y$.
In experimental study design parlance, $X$ is the \emph{treatment}\footnote{
  The term \emph{treatment} comes from the practice of randomized controlled clinical trials,
  where it strictly denotes a binary variable (treatment vs.\ control group) that is randomly assigned.
  In statistical parlance, however, the term is routinely applied more generally
  to indicate any variable
  whose effect on an outcome is being analyzed.
  Under this general meaning,
  a treatment variable may not be randomly assigned~\cite{treatment-nonrandom}
  (for example, in an observational study)
  or it may be continuous~\cite{treatment-continuous} instead of binary.
  In this paper, we use ``treatment'' with this relaxed, more general meaning---as a synonym of ``exposure''.
}
or \emph{exposure},
and $Y$ is the \emph{outcome} or \emph{response}.
Here is a concrete example in the domain of software engineering,
which we will look into more closely in \autoref{sec:intro-confounding}:
estimating how using a different programming language (treatment $X$)
affects the quality (outcome $Y$) of a program implemented in that language.
If we have observational data about $X$ and $Y$,
we may simply fit a statistical model---anything from a simple linear regression to fancier devices---on these data,
and interpret the fitted model's parameters
connecting $X$ and $Y$ as an estimate of this effect.
In this process,
a well-known snag is \emph{omitted variable} bias:
if there exists another \emph{unmeasured} variable $Z$
that affects both $X$ and $Y$,
our estimate of the effect $X \to Y$ will spuriously include
the combined effect of $X$ and $Z$ instead of the effect of $X$ alone.
Continuing our empirical software engineering example,
$Z$ may be a programmer's intrinsic skills,
which are all but sure to also affect the quality $Y$ of the written programs.
In such a scenario,
additional data about $X$ and $Y$ (i.e., sampling more datapoints)
is not going to help;
in fact, it may just entrench our reliance on the biased estimate
by reducing its variance and giving the false impression of reliability or ``significance''.

Omitted variable bias is a widespread risk of any statistical analysis
of observational data,
regardless of whether one employs
frequentist~\citep{holtBT14state,grenTF17group,torkar22stats}, Bayesian~\citep{levenBBT2024bwt,penzenstadlerTM2022breath,furiaTF22guidelines},
or other kinds of machine learning models~\citep{Andaur21biasML, mehrabi22surveyML}.
In fact,
it is always possible that \emph{some} relevant variable
was not measured, because it was
unknown, inaccessible, or impractical, time-consuming, or expensive
to measure with reasonable accuracy.
This paper's main contribution is presenting several \emph{mitigation strategies}
to cope with omitted variable bias,
and demonstrating them in scenarios and examples that are
relevant for software engineering empirical data analysis.

Of course, the ideal approach to avoid omitted variable bias is
running a fully \emph{controlled experiment},
where the treatment $X$ is assigned randomly,
and the corresponding values of $Y$ are recorded.
Controlled experiments are the gold standard in science
precisely because they protect from omitted variable bias
even when we don't even know which other unmeasured variables may
bias our estimate---since randomly assigning the treatment effectively
removes its dependencies on any (unmeasured) confounder.
The obvious reason why controlled experiments are not more common
is that they are generally very expensive to run.
In a field like empirical software engineering,
proper controlled experiments are prohibitively challenging to design and run
on the time scale of real-world software development---as opposed to ``toy'' programming tasks.
In contrast, there is abundant observational data
from software repositories that span large systems developed
over years by many developers;
but discovering genuine causal effects using
these purely observational data must contend with omitted variable bias.
Therefore, this paper focuses on
mitigation strategies for omitted variable bias
when analyzing observational data.

Omitted variable bias 
is a common and relevant problem---especially for empirical software data.
However, it is by no means the only pitfall of analyzing empirical data:
as we further discuss in \autoref{sec:bg},
there are plenty of other challenges
such as the included variable bias~\citep{furiaTF23causal},
``precisely inaccurate'' analyses that hide biases~\citep{mcfarlandM15inaccurate},
and unrepresentative population samples~\citep{bradleyKISMF2021unrep}.
While each challenge requires different measures,
they all likely involve trade-offs between costs and benefits---similarly to the present paper's outlook on dealing with omitted variable bias.

As we discuss more broadly in \autoref{sec:related},
there has been a growing interest in
adopting robust, modern statistical practices in
empirical software engineering research---especially those
based on causal analysis.
This paper contributes to this line of research,
focusing on the practical and widespread problem of omitted variable bias.

\subsection{Contributions}
\label{sec:contributions}

This paper makes the following contributions:

\begin{itemize}
\item Demonstrates the significance of omitted variable bias
  when analyzing empirical software engineering (observational) data;

\item Presents statistical methods
  to detect, quantify, and mitigate omitted variable bias
  when analyzing empirical data;

\item Provides simple guidelines for empirical software engineering researchers
  to apply those mitigation strategies in practical settings;

\item For reproducibility, all data and analysis scripts are available online:
    \begin{center}
        {\small \textsc{replication package:}}$\quad$ \url{https://figshare.com/s/fe607d8eb7c4cedbac75}$\quad$ \cite{omitted-vars-replication}
      \end{center}

\end{itemize}

The main aims of this paper are \emph{methodological}:
the illustrative examples of \autoref{sec:intro-confounding}
and \autoref{sec:teamwork} are based on real data,
but are not meant to lead to novel findings about their domains.
Instead, they demonstrate plausible scenarios where
confounding may occur, and explore different mitigation strategies.

\subsection{Scope and Limitations}
\label{sec:scope-limitations}
This paper focuses on a specific, yet widespread, challenge
in empirical software engineering:
estimating and mitigating omitted variable bias
when analyzing observational data.
The techniques we illustrate use
structural causal models (DAGs, or directed acyclic graphs)
to model the possible underlying causal relations between variables;
adjustment sets to determine which variables should be included
in a statistical model to make its inferences consistent with the causal relations;
and sensitivity analysis to systematically explore
the robustness and validity of the inferred results
relative to a range of causal assumptions.
The methods we present
are applicable both \emph{a posteriori}---to analyze non-experimental data---and \emph{a priori}---to inform the design of observational or quasi-experimental studies
and to proactively outline the boundary of their validity.

These methods provide an accessible and practical framework
for dealing with a common source of confounding in
observational studies.
However, they do not cover all confounding problems
and are not suitable for every kind of empirical study design:
\begin{itemize}
\item We focus on recovering \emph{causal} effects,
  as opposed to optimizing
  predictive performance or model succinctness in a purely statistical setting;
  in related work, we demonstrated scenarios where these goals
  lead to different results~\citep{furiaTF23causal}.

\item The methods we present in the paper are well-suited
  for \emph{cross-sectional} studies, where data from a population
  is collected at a specific point in time.
  These studies are very common in empirical software engineering.
  For comparison, \autoref{sec:longitudinal}
  outlines a few approaches that extend confounding analysis
  to time-dependent data,
  which arise in longitudinal or panel studies.

\item Omitted variable bias is only one of many different threats
  to the validity of empirical studies.
  Different kinds of confounding may require different statistical approaches~\citep{mcelreath2020statistical};
  more generally, other sources of bias
  (selection bias, measurement bias, \ldots)
  may jeopardize a study's design validity
  at different levels~\cite{experiments-book}.

\item The DAG-based methods we consider in this paper
  have been gaining traction, as they support rigorous, yet intuitive
  foundations for causal modeling, and they are sufficiently general
  to accommodate a wide range of applications.
  However, other causal analysis frameworks exist---such as difference-in-differences,
  instrumental variables,
  regression discontinuity designs,
  matching methods, and synthetic control---that may be more suitable in certain contexts.
  Without going into details,
  a key difference of these approaches to causality
  is that they usually address confounding
  through preprocessing, design, or data structure
  (for example, relying on timing or instruments)
  rather than by modeling and adjusting within a statistical model
  as in DAG-based approaches~\citep{Stuart10MM}.
  A more detailed presentation of these alternatives
  is outside this paper's scope,
  and is available elsewhere~\citep{ImbensJ09dev,AtheyI17SOA}.

\item The DAG-based methods we demonstrate in the paper
  rely on several technical assumptions including: the
  absence of causal loops (acyclicity of the DAG),
  probabilistic relations between variables
  (as opposed to purely deterministic relations,
  which can only be modeled indirectly~\cite{det-vars}),
  and enough evidence to inform a plausible
  causal structural model.
  \autoref{sec:bg} presents the underlying methods in detail,
  compares them to other, related approaches,
  and outlines their relevance for empirical software engineering research.
\end{itemize}

\subsection{Organization}

\autoref{sec:bg} first presents relevant related work
on the topics of mitigating biases in statistical analysis and, more generally, best statistical practices for empirical software engineering; then, it introduces the key notations and concepts that the rest of the paper relies on.
The rest of the paper demonstrates how to deal with omitted variable bias in two illustrative examples: \autoref{sec:1stcase} uses a comparatively simpler model to estimate the effect of programming languages on code quality (also mentioned in the introduction), which serves as a relatively uncomplicated scenario. Then, \autoref{sec:2ndcase} uses a more complex model to analyze the relation between team size (how many developers work on a project) and effort (how long it takes to complete the project). Although the two illustrative examples differ in complexity, they are both based on realistic scenarios and models that we investigated in our previous work~\citep{furiaTF22guidelines,feldtSFT23sysrev}.
\autoref{sec:disc} serves as a high-level summary of the whole article,
presenting \emph{guidelines}, in the form of a sequence of analysis steps
that generate fundamental information to support the design of a study
that mitigates omitted variable bias.
Finally, \autoref{sec:concl} concludes the paper with a short summary of the main contributions.

\section{Related Work and Background}\label{sec:bg}

This section starts (in \autoref{sec:related}) with a brief overview
of the origins of causal analysis techniques for observational data,
and how they have been adopted in empirical sciences
(including software engineering).
Then, \autoref{sec:dags} and \autoref{sec:confounding}
introduce the key concepts (causal relations, DAGs, confounders)
of causal analysis that we will develop in the rest of the paper.
Finally, \autoref{sec:other-confounding}
positions the paper's contributions by relating them to
other forms of confounding and causal inference bias.

\subsection{Background}
\label{sec:related}

One of science's ultimate goals is
understanding the processes that underlie observed phenomena.
This means discovering \emph{cause/effect} relations between variables,
as opposed to mere statistical \emph{associations}.
While the roots of causal analysis date back to Neyman's
potential outcomes framework~\citep{neyman23thesis},
a comprehensive understanding of causality has emerged only later,
in the late part of the 20th century.
The key milestones in the development of a robust understanding
of causality include:
\begin{enumerate*}
\item \citet{rubin1974estimating} built upon Neyman's pioneering work,
  introducing a framework for causality in nonrandomized (observational) studies.
\item \citet{angristIR96iv} introduced instrumental
  variables\footnote{ In a nutshell, an instrumental variable is a
    variable that acts like a natural experiment on the treatment. }
  based on the Neyman-Rubin framework.
\item Working at about the same time as Rubin, Pearl
  started focusing on structural (graph) models~\citep{pearl82bayes},
  which culminated in his celebrated techniques for
  rigorously analyzing causal effects and confounding~\citet{pearl09a@reason}.
\end{enumerate*}

Among Pearl's work, directed acyclic graphs (DAGs)
have become widely used
to model the structural dependencies between observed (and unobserved) variables.
As we will demonstrate already with the simple example of \autoref{sec:dags},
DAGs are, first of all, a practical notation to specify
causal relations.\footnote{
  An alternative approach to modeling causal relations is
  structural equation modeling (SEM).
  Semantically, Pearl’s causal DAG framework
  and SEM with latent variables are closely related.
  Specifically, Pearl’s original work built on SEM,
  and modern presentations of the causal DAG framework
  explicitly connect it to SEM~\citep{Bollen2013}.
  Methodologically, however, the two approaches differ in their focus:
  causal DAGs (especially as we use them in the paper)
  specialize in modeling structural causal relations
  among (observed or latent) variables;
  SEM, instead, tends to mix measurement aspects
  (e.g., the relation between latent constructs and indicators)
  and structural ones (e.g., the relation between different constructs).
}
They also support techniques to \emph{estimate}
the strength of the causal relation among some nodes in the graph.
Usually this is done by constructing a (linear) statistical model
among variables, selected according to the DAG's structure.
(We will demonstrate this shortly in \autoref{sec:confounding}.)
Nowadays, causal analysis based on DAGs is
routinely used in disciplines with a strong empirical component
such as
medicine~\citep{dag-paediatrics,dag-health}, epidemiology~\citep{dag-epidemiology},
economics~\citep{dag-economics}, and biology~\citep{dag-biology}.

A key question when working with DAGs is
how to build a realistic DAG in the first place.
In fields where the underlying fundamental mechanisms
are well understood, a DAG can be built
based on expert knowledge and previous work.
Another approach is \emph{causal discovery}
(also called structural learning),
which tries to identify causal relations from data~\citep{Spirtes:2016causal}.
While causal discovery algorithms have made significant progress
in the last decade~\citep{causal-discovery-survey}---in part on the wave of the recent machine learning boom---they remain \emph{heuristic} approaches that work correctly only under
precise assumptions about the possible interactions.
This limitation is intrinsic,
as one of the fundamental results of Pearl's framework
is that causal relations cannot be inferred from data alone
(at least not without limiting assumptions).
Regardless of whether they come from expert knowledge,
are inferred heuristically from data, or
simply encode some (plausible) hypotheses,
DAGs remain a practical tool
to precisely denote, validate, and reason about
the causal relations in a system.

While causal analysis techniques
are not widely used in software engineering empirical research,
they are gradually gaining traction.
\citet{Siebert:2021causal}'s recent survey
reports 31 studies in empirical software engineering
that targeted some kind of causal analysis technique.
All of the reviewed papers were published in the last 15 years,
which confirms that causal analysis is not yet an established practice
but is slowly gaining popularity.
The majority (17 out of 31) of papers
reviewed by \citeauthor{Siebert:2021causal}
are about fault localization
and refer to \citet{baah2010causal}---the first contribution that tried to
apply a causal view instead of the traditional, purely correlational
analysis that is commonplace in fault localization techniques~\cite{fl-survey,Zou:2021,RF-EMSE24-FauxPy}.
Testing is another area targeted by several of the studies reviewed by \citeauthor{Siebert:2021causal};
these include applications to mutation testing~\cite{lee2021causal},
simulation testing~\citep{clark2022testing}, and A/B testing~\citep{liu2022bayesian}.
The other papers reviewed by \citeauthor{Siebert:2021causal}
target various topics such as performance analysis
(in one case still linked to fault localization~\citep{scholz2021empirical}).

In the last few years,
some more empirical software engineering research was published,
targeting varied topics such as:
\begin{enumerate*}
\item modeling rules of human knowledge and how they are made available to
  artificial intelligence systems~\citep{heyn2022structural};
\item studying the impact of social media posts on the popularity of open-source
  projects~\citep{fang2022slick};
\item analyzing dependencies in configurable software systems~\citep{halpern2015modification};
\item studying the impact of programming languages on coding competitions~\citep{furiaTF23causal}.
\end{enumerate*}

More broadly, these techniques belong to a broad and growing body of work on modernizing statistical practices~\citep{mcelreath2020statistical}.
This work has been the response to the realization
that some ``traditional'' and widely used statistical methods
have limitations in terms of interpretability and real-world significance~\cite{ASA-statement,riseup,abandon-significance}.
While improving statistical practices is only a piece of the puzzle,
inadequate methods have certainly concurred
to the issues with replicability
experienced in empirical research areas as diverse as medicine~\cite{all-false}, psychology~\cite{reprod-psych}, and economics~\cite{reprod-econ}.
Since empirical software engineering research
shares challenges and practices with these domains,
adopting more robust data analysis practices
will benefit 
the quality and impact of our research field~\cite{repr-msr,repr-pl,repr-empirical,SmellsAnalytics,repr-se}.

In conclusion, there is growing interest in
understanding the concepts of causal analysis
and applying them to analyzing software engineering data.
The present paper further supports this trend
by demonstrating how the framework of causal analysis
can help mitigate the pervasive issue of omitted variable bias.

\begin{figure}[!tb]
  \centering
\begin{subfigure}{0.26\linewidth}
    \centering
    \begin{equation*}
      \begin{aligned}
        X &\sim \dist{Normal}(0, 1) \\
        \epsilon &\sim \dist{Normal}(0, 1) \\
        Y &= b X + \epsilon \\
        \phantom{X} 
      \end{aligned}
    \end{equation*}
    \begin{tikzpicture}[->,very thick]
      \node[var] (X) {$X$};
      \node[var,right=of X] (Y) {$Y$};
      \draw (X) -- (Y) coordinate[midway] (M);
      \phantom{\node[var,above=of M] (Z) {$Z$};}
    \end{tikzpicture}
    \caption{Process $p_1$:
      $Y$ depends on $X$, with some measurement error~$\epsilon$.}
    \label{fig:XY}
  \end{subfigure}
\hfill
  \begin{subfigure}{0.29\linewidth}
    \centering
    \begin{equation*}
      \begin{aligned}
        X &\sim \dist{Normal}(0, 1) \\
        Z &\sim \dist{Normal}(0, 1) \\
        \epsilon &\sim \dist{Normal}(0, 1) \\
        Y &= b X + c Z + \epsilon
      \end{aligned}
    \end{equation*}
    \begin{tikzpicture}[->,very thick]
      \node[var] (X) {$X$};
      \node[var,right=of X] (Y) {$Y$};
      \draw (X) -- (Y) coordinate[midway] (M);
      \node[var,above=of M] (Z) {$Z$};
      \draw (Z) -- (Y);
    \end{tikzpicture}
    \caption{Process $p_2$:
      $Y$ depends on $X$ and $Z$, with some measurement error~$\epsilon$.}
    \label{fig:XYZ}
  \end{subfigure}
\hfill
  \begin{subfigure}{0.3\linewidth}
    \centering
    \begin{equation*}
      \begin{aligned}
        Z &\sim \dist{Normal}(0, 1) \\
        X &= d Z + \epsilon_X \\
        Y &= b X + c Z + \epsilon_Y \\
        \epsilon_{X,Y} &\sim \dist{Normal}(0, 1)
      \end{aligned}
    \end{equation*}
    \begin{tikzpicture}[->,very thick]
      \node[var] (X) {$X$};
      \node[var,right=of X] (Y) {$Y$};
      \draw (X) -- (Y) coordinate[midway] (M);
      \node[var,above=of M] (Z) {$Z$};
      \draw (Z) -- (Y);
      \draw (Z) -- (X);
    \end{tikzpicture}
    \caption{Process $p_3$:
      $Y$ depends on $X$ and $Z$, and $X$ also depends on $Z$,
      with some measurement error $\epsilon$.}
    \label{fig:XYZX}
  \end{subfigure}
  \caption{Three possible processes where variable $Y$
    depends on variables $X$ and $Z$,
    and the corresponding DAGs capturing these structural relations.}
  \label{fig:XY-gen}
\end{figure}

\begin{figure}[!tb]
  \centering
\begin{subfigure}{0.35\linewidth}
    \centering
    \begin{equation*}
      \begin{aligned}
        Y_i &\sim \dist{Normal}(\mu_i, \sigma) \\
        \mu_i &= \alpha + \beta X_i
      \end{aligned}
    \end{equation*}
    \caption{Model $m_1$: $Y$ is conditioned on $X$ only.}
    \label{fig:mX}
  \end{subfigure}
\hspace{20mm}
  \begin{subfigure}{0.4\linewidth}
    \centering
    \begin{equation*}
      \begin{aligned}
        Y_i &\sim \dist{Normal}(\mu_i, \sigma) \\
        \mu_i &= \alpha + \beta X_i + \gamma Z_i
      \end{aligned}
    \end{equation*}
    \caption{Model $m_2$: $Y$ is conditioned on both $X$ and $Z$.}
    \label{fig:mXZ}
  \end{subfigure}
  \caption{Two linear regression models that capture the dependence between
    $X$, $Y$, and $Z$.}
  \label{fig:XYZ-regression}
\end{figure}

\subsection{Causal Dependencies and DAGs}
\label{sec:dags}

Let's go back to the example of two observed variables $X$ and $Y$, which we briefly introduced in \autoref{sec:intro}.
Imagine that the process that determines the values of $X$ and $Y$ is
perfectly known.
In turn, we consider
each of the three processes described by the equations in \autoref{fig:XY-gen}.
For simplicity, all our examples use \emph{linear} dependencies
and normal distributions,
but the same line of thought is applicable
to more complex, non-linear dependencies.

Process $p_1$ in \autoref{fig:XY}
produces
values of $X$ that are drawn randomly from a normal distribution
with zero mean and unit standard deviation;
and values of $Y$ that are linearly proportional to those of $X$.
Even in such an ideal scenario, any empirical \emph{measure}
of $X$ and $Y$ will include some measurement error $\epsilon$,
which process $p_1$ models as another normal random variable
that, together with $X$, determines the value of $Y$.
We stress that we interpret \autoref{fig:XY}'s equations
as capturing the causal dependencies among variables:
$X$ and $\epsilon$ are drawn randomly (and independent of each other),
and their random values
determine (``cause'') the value of $Y$ in each draw.
Correspondingly, the DAG (directed acyclic graph~\citep{pearl09a@reason})
in \autoref{fig:XY}
captures this causal relation between $X$ and $Y$ in a qualitative way:
an edge connects $X$ to $Y$ to denote
that the values of $X$ and $Y$ are related;
furthermore,
the edge is directed from $X$ to $Y$
to denote that changing $X$ directly affects $Y$,
that is, it causes $Y$ to change.

Process $p_2$ in \autoref{fig:XYZ}
involves a third variable $Z$.
Just like variable $X$,
$Z$'s values are drawn randomly from a normal distribution
with zero mean and unit standard deviation.
Then, 
the values of $Y$ are
a linear combination of $X$ and $Z$---still with a term $\epsilon$ to account for measurement errors.
Variables $X$ and $Z$ are independent of each other;
the DAG in \autoref{fig:XYZ}
clearly shows this independence,
since it does not include any edge between $X$ and $Z$.

Process $p_3$ in \autoref{fig:XYZX}
still involves the three variable $X$, $Y$, and $Z$.
Now, $Z$ is the only variable that is drawn independently;
in contrast, $X$ depends linearly on $Z$,
and $Y$ depends linearly on a combination of $X$ and $Z$.
As usual, the DAG in \autoref{fig:XYZX}
visualizes these relations among variables,
showing, in particular, that there is both a direct relation
between $X$ and $Y$ (edge $X \to Y$)
and an indirect relation through $Z$
(path $\underrightarrow{X \leftarrow Z \to Y}$).

\subsection{Inference and Confounders}
\label{sec:confounding}

Consider an empirical study
whose goal is to \emph{estimate}, from a sample of the data,
the parameters of a statistical model
that captures the relations among observed variables.
Let's shift perspective on our illustrative examples:
now, we are given a data sample $D_1$, $D_2$, $D_3$
respectively produced by each process $p_1$, $p_2$, $p_3$.
Each data sample consists of many triples $(x_i, y_i, z_i)$
of concrete values taken by $X$, $Y$, and $Z$\footnote{Process $p_1$'s data sample only consists of pairs,
  since this process does not include a variable $Z$.}
in the $i$th observation---that is, the process's $i$th draw.
Now, let's assume that 
we do not know the equations governing the generation process,
but we want to quantitatively estimate 
the relation between $X$ and $Y$ from the observed sample.

Given that we are dealing with linear relations and normal distributions,
we will use a \emph{regression model} to fit the data.
A key choice is whether to include variable $Z$ in the model:
regression model $m_1$, shown in \autoref{fig:mX}, ignores $Z$,
whereas model $m_2$, shown in \autoref{fig:mXZ}, includes $Z$ as a predictor.
In both cases, after fitting a model on the data,
the value of parameter $\beta$
will be the model's estimate of the corresponding $b$ in \autoref{fig:XY-gen}---in other words, $\beta$ estimates the causal ``effect'' of
the predictor $X$ on the outcome $Y$.
\autoref{tab:confounded-or-not}
shows the result of this exercise,
when \autoref{fig:XY-gen}'s processes
use concrete values $b = 0.4$, $c = 0.7$, and $d = 0.2$
for their parameters.

\begin{table}[!tb]
  \centering
\begin{subtable}{0.24\linewidth}
  \begin{tabularx}{\linewidth}{c|ccc}
    & \multicolumn{3}{c}{\textsc{process}} \\
    \textsc{model} & $p_1$ & $p_2$ & $p_3$ \\
    \midrule
    $m_1$ & \suppOK & \suppOK & \suppNO
    \\
    $m_2$ &  & \suppOK & \suppOK 
  \end{tabularx}
  \caption{Whether each \textsc{model} $m$
    estimates correctly (\suppOK) or incorrectly (\suppNO)
    the effect $b$ of $X$ on $Y$ in \textsc{process} $p$.}
  \label{tab:confounded-bool}
\end{subtable}
\hspace{10mm}
\begin{subtable}{0.632\linewidth}
  \begin{tabularx}{\linewidth}{c|rrr rrr rrr}
    & \multicolumn{3}{c}{$\beta$}
    & \multicolumn{3}{c}{$\gamma$}
    & \multicolumn{3}{c}{$\sigma$}
    \\
    \cmidrule(lr){2-4}
    \cmidrule(lr){5-7}
    \cmidrule(lr){8-10}
    \textsc{model}
    & \multicolumn{1}{c}{$p_1$}
    & \multicolumn{1}{c}{$p_2$}
    & \multicolumn{1}{c}{$p_3$}
    & \multicolumn{1}{c}{$p_1$}
    & \multicolumn{1}{c}{$p_2$}
    & \multicolumn{1}{c}{$p_3$}
    & \multicolumn{1}{c}{$p_1$}
    & \multicolumn{1}{c}{$p_2$}
    & \multicolumn{1}{c}{$p_3$}
    \\
    \midrule
    $m_1$
    & 0.40  & 0.40 & 0.54
    &       &      &      
    & 1.00  & 1.22 & 1.22
    \\
    $m_2$
    &       & 0.40 & 0.40
    &       & 0.70 & 0.70
    &       & 1.00 & 1.00
  \end{tabularx}
  \caption{For each process $p_1$, $p_2$, $p_3$
    the values of parameters $\beta$, $\gamma$, $\sigma$
    in \textsc{model} $m_1$ or $m_2$
    fitted on data generated by the process.
    In these experiments, the parameters of \autoref{fig:XY-gen}'s
    processes are set to $b = 0.4$, $c = 0.7$, $d = 0.2$.}
  \label{tab:fitted-values}
\end{subtable}
\caption{Estimating the parameters
  of \autoref{fig:XY-gen}'s processes $p_1$, $p_2$, and $p_3$
  with \autoref{fig:XYZ-regression}'s regression models $m_1$ and $m_2$.}
  \label{tab:confounded-or-not}
\end{table}

\paragraph{Process $p_1$.}
The case of process $p_1$ is unproblematic:
since the process does not involve any variable other than $X$ and $Y$,
regression model $m_1$ accurately infers the value of parameter $\beta \simeq 0.40 = b$,
reflecting the true dependency between $X$ and $Y$.
The fitted regression model also accurately infers
the standard deviation $\sigma = 1.00$ of the error term $\epsilon$.
Obviously, model $m_2$ is inapplicable to analyze data produced by $p_1$,
since this includes no variable $Z$.

\paragraph{Process $p_2$.}
The case of process $p_2$ is more interesting,
since we may analyze its data using either model $m_1$ or model $m_2$.
As we expect,
regression model $m_2$ accurately infers the value of parameter $\beta = b = 0.40$,
again reflecting the true dependency between $X$ and $Y$.
Somewhat less obviously,
the simpler model $m_1$
still infers the same correct value of parameter $\beta = b = 0.40$,
even if $Y$ also depends on $Z$ in generation process $p_2$.
With model $m_1$, the effect of $Z$ on $Y$ has spilled
into the estimate of the standard deviation $\sigma$ of the error term,
which is in fact equal to $1.22$, 
greater than the ``true'' error standard deviation $1.0$.\footnote{
  Equivalently, we can rewrite \autoref{fig:XYZ}'s generative equations
  as $Y = bX + E$, where $E \sim \mathrm{Normal}(0, \sqrt{1 + c^2})$;
  if $c = 0.7$, $\sqrt{1 + c^2} \simeq 1.22$,
  which is exactly the inferred value of $\sigma$
  in $m_1$ fitted on data from $p_2$.
}

\paragraph{Process $p_3$.}
Regression model $m_1$ cannot accurately account
for the more intricate data dependencies of process $p_3$.
In fact, it overestimates the effect $\beta = 0.54 > 0.40 = b$
of $X$ on $Y$;
now, this effect also includes the ``spurious'' correlation
introduced by $Z$, which simultaneously affects $X$ and $Y$---as \autoref{fig:XYZX}'s DAG clearly shows.
Variable $Z$ is called a \emph{confounder},
since it mixes up the true effect of $X$ on $Y$
(path $X \to Y$ in the DAG)
with an indirect, spurious correlation
(path $\underrightarrow{X \leftarrow Z \to Y}$ in the DAG)
that does not correspond to an actual data dependency
but is just a figment of using an inadequate statistical model.
Model $m_2$ makes up for $m_1$'s shortcomings
by including $Z$ among its predictors;
this is enough to cancel $Z$'s confounding effect
on the link between $X$ and $Y$.
Indeed, \autoref{tab:fitted-values} shows that
all parameters---crucially, the effect $\beta = 0.40 = b$ of $X$ on $Y$---are correctly estimated.

\paragraph{}
Using a model such as $m_1$
to analyze data from a process whose dependencies include a confounder
(like process $p_3$)
is an instance of \emph{omitted variable bias}.
The rest of the paper describes more realistic illustrative examples
where omitted variable bias may occur,
and presents various mitigation strategies to counter the bias
and recover a precise estimate of the causal effect linking
treatment $X$ and outcome $Y$.

From the toy examples of this section,
we can start to glean how omitted variable bias is commonplace
in realistic settings.
Even when our empirical data are rich and include
many different variables,
there is always a chance that we are missing some \emph{other}
variables that, like $Z$, confounds the effect of interest.
Even if we are aware of possible confounders,
measuring them to include them in our model
(like model $m_2$ does)
may be expensive,
impractical, or impossible.
For example, the confounder may lack a good operationalization,
or it may be inaccessible because its values were not recorded
and the process is not repeatable.
These observations motivate the main contributions of the paper,
which demonstrate how to identify and mitigate the wicked effects
of confounders in a variety of practical scenarios.

\paragraph{Correction and sensitivity analysis.}
\label{sec:sensitivity-methodologically}
Broadly speaking,
the rest of the paper illustrates two complementary approaches
to deal with omitted variable bias when analyzing observational data.
First, we will present techniques that can \emph{correct}
for possible omitted variable bias.
Based on structural models of causality (see \autoref{fig:XY-gen}'s DAGs),
such techniques suggest how to model the data in a way
that filters out confounding effects and recovers more precisely
the strength of actual causal relationships.

As we will see in our examples,
correcting for possible confounders is not always possible or practical.
On the one hand, some confounding variables may simply be inaccessible;
on the other hand, the model of causality that underlies the application
of adjustment techniques may itself be imprecise or uncertain.
In such cases, a \emph{sensitivity analysis} can
mitigate uncertainty and clarify the boundary of validity of our findings.
In a nutshell, a sensitivity analysis is a sort of ``what if''
analysis that quantifies the robustness of the results
under different assumptions about the underlying process.\footnote{
It is important to notice that, despite their superficial similarities,
  the practice of sensitivity analysis and the malpractice of data dredging
  (also known as ``$p$-hacking'') have opposite goals:
  ``In $p$-hacking, many justifiable analyses are tried, and the one
  that attains statistical significance is reported. In sensitivity
  analysis, many justifiable analyses are tried, and all of them are
  described.'' \citep[p.~319]{mcelreath2020statistical}.
}
A sensitivity analysis can corroborate results
and expose limits of validity, but is not expected to give
any kind of definitive, binary answer.

\subsection{Other Forms of Confounding}
\label{sec:other-confounding}

The term ``confounding'' is used to denote
different, related concepts in the statistical analysis of empirical data~\citep{greenland01confounding}.
In this paper, we use ``confounding''
to denote bias in the estimate of a \emph{causal effect}---as
we demonstrated in a nutshell in the previous sections.
This notion of confounding is customary in modern causal analysis~\citep{pearl09a@reason},
and in related approaches to mitigate confounding, such as instrumental variables~\citep{angristIR96iv}.

An early usage of ``confounding'' in statistics was to
denote \emph{noncollapsibility}~\citep{greenland01confounding,yule03bio}.
In a nutshell,
the effect of $X$ over $Y$ is \emph{collapsible over $Z$}
if the marginal association between $X$ and $Y$
(obtained by averaging over $Z$)
is the same as their conditional association
(obtained by conditioning on $Z$);
noncollapsibility holds when the two associations differ.
While noncollapsibility is a purely statistical notion,
the notion of confounding is rooted in causal relations,
which depend not only on data but also on the data generation process~\citep{pearl09a@reason};
therefore, the two concepts overlap but are distinct.
In particular,
noncollapsibility may signal causal confounding
(for example, in a scenario like \autoref{fig:XYZX});
but it may also arise from nonlinearities without confounding.
For example,
consider a scenario with the same causal structure $X \to Y \leftarrow Z$ as in \autoref{fig:XYZ}
but where $Y$ has a non-linear logistic relation with the linear combination
of $X$ and $Z$.
Since the causal structure is the same as in \autoref{fig:XYZ}
there is no confounding;
however, the effect of $X$ on $Y$ depends on whether we condition on $Z$,
and hence there is noncollapsibility.

In classical frequentist statistics,
the term ``confounding''
denotes an experimental design that makes
two effects indistinguishable from data~\citep{fisher35exp}.
This usage of the term refers to statistical identifiability,
and does not have, in general,
any relation to causal effects.
For example, a so-called \emph{fractional} factorial design
omits some combinations of factor levels, which may
introduce confounding in this sense.

In fields such as psychometrics,
confounding is often characterized as a measurement problem
or, more generally, an issue of \emph{experimental design}~\citep{cook1979quasi}.
In the present paper, in contrast, we take the main point of view
of analyzing observational data,
having little or no control on the process
that generated the data.

While all these senses of the term ``confounding''
are related,
the causal meaning that we follow in this paper
captures a key challenge 
the analysis of observational data,
and provides the clearest characterization of the omitted variable bias.

\subsection{Time-dependent Data}
\label{sec:longitudinal}

As we explained in \autoref{sec:scope-limitations},
the techniques presented in this paper are not
directly applicable to analyze confounding
in \emph{time-dependent} data,
such as those that are collected in longitudinal studies.
This section is a brief detour into approaches
that take time into account.

As a concrete example, consider studying
the impact of code reviews on the
quality of a software project.
A longitudinal study would consider a series of
interventions $R_1, R_2, \ldots, R_n$,
where each $R_k$ denotes whether release $k$
of a project was ($R_k = 1$) or wasn't ($R_k = 0$)
reviewed before being merged into the main branch.
The study's goal is to determine the cumulative effect
of the interventions $R_k$ on the project quality $Q$
at release $n$.
Covariates, possibly confounders, may also be time dependent,
such as for the \emph{experience} $E_k$ of the developer
reviewing release~$k$.

Even though it is common for the observational data collected
in software engineering studies (in particular by mining software repositories)
to have a temporal dimension,
a study may choose to abstract away the temporal dependencies and
perform a cross-sectional analysis---possibly as a stepping stone towards
a full-fledged longitudinal analysis.
In the aforementioned domain of code reviews,
a cross-sectional study would simply consider each data point
as a ``snapshot'' of the relations $R_k \to Q_k$
at release (time) $k$.
\citet{ImaiKim2019fixed}
elucidate the impact of abstracting away time in longitudinal data,
and under what assumptions it does not introduce confounding.

On the other hand, there have been several proposals in the literature
of techniques to model and correct for confounding
in time-dependent data.
Marginal structural models (MSMs)~\cite{Robinsetal2000MSM}
and structural nested models~\cite{VansteelandtJoffe2014SNM}
are especially interesting from our perspective,
since they are built atop the same causal DAGs that
underlie this paper's technique.
For instance, MSMs
build DAGs that represent the relation between variables
at different time steps.
In the code review domain, for example,
we could posit a causal relation $E_k \to E_{k+1}$
if a reviewer may opt in to be assigned to the next code review.

\section{Confounding in
  Programming Languages and Code Quality}
\label{sec:intro-confounding}
\label{sec:1stcase}

Our first illustrative example follows closely the fundamental structure of
\autoref{sec:confounding}'s prototypical example,
while recasting it in a more realistic setting.
Our goal is estimating the effect of
using different programming languages (predictor variable $\var{Language}$)
on the quality of the produced code written in that language
(outcome variable $\var{Quality}$).

\begin{figure}
  \centering
  \begin{tikzpicture}[->,very thick]
    \node[var] (X) {\var{Language}};
    \node[var,right=of X] (Y) {\var{Quality}};
    \draw (X) -- (Y) coordinate[midway] (M);
    \node[var,above=of M] (Z) {\var{Skill}};
    \draw (Z) -- (Y);
    \draw (Z) -- (X);
  \end{tikzpicture}
    \caption{The effect of \var{Language} on code \var{Quality} is confounded by the programmer's \var{Skill}.}
    \label{fig:language-quality-skill}
\end{figure}

As we discussed in depth in related work~\citep{furiaTF22guidelines},\footnote{
  In this section, we analyze a subset of the same data
  collected by others~\citep{FSE,TOPLAS}
  that we also analyzed in our previous work~\citep{furiaTF22guidelines}.
  This section's and our previous work's analysis~\citep{furiaTF22guidelines}
  are, however, orthogonal:
  in this section, we analyze possible confounding effects based on
  causal assumptions;
  in contrast,
  \citep{furiaTF22guidelines}'s analysis is purely statistical
  and targets Bayesian statistical modeling practices.
  }
a programmer's ability
(expressed by variable $\var{Skill}$)
is likely to confound the causal effect of $\var{Language}$ on $\var{Quality}$.\footnote{
  It's plausible several other confounders of this causal relation exist~\citep{furiaTF22guidelines}; for clarity of presentation,
  we only consider \var{Skills}, as if it captures the effects of
  other possible confounders of the same relation.
}
Namely, a more skilled programmer is likely to produce
higher-quality code ($\var{Skill} \to \var{Quality}$);
and programmers with different abilities
may prefer to work with certain programming languages over others
($\var{Skill} \to \var{Language}$).
\autoref{fig:language-quality-skill}
summarizes these relations by means of a DAG,
which is isomorphic\footnote{
  \autoref{fig:language-quality-skill} and
  \autoref{fig:XYZX} are isomorphic as graphs.
  However, the variables associated with corresponding nodes
  may have different characteristics---most notably, $X$ is continuous
  whereas \var{Language} is categorical.
}
to \autoref{fig:XYZX}'s abstract DAG.

As discussed in \autoref{sec:confounding},
this DAG structure entails
that, if we want to estimate the true effect of
\var{Language} on \var{Quality} for observational data---where we cannot control the effect of \var{Skill} on \var{Language},
that is, we cannot randomize which language each programmer will use---we need to also \emph{condition} on the confounder \var{Skill}.
In practice, this may be impossible because precisely
measuring \var{Skill} is not easy:
for example, if the data comes from a source code repository,
the identity of the programmers may be unknown;
even if we have access to a programmer's identity,
it may be practically cumbersome to reliably assess their programming skills.

The rest of this section demonstrates how we can mitigate
the effects of this unobserved variable bias,
in a way that we can still get something out of our
observational data about languages and code quality---even in such a tainted scenario.

\subsection{Data: Programming Languages and Code Quality}
\label{sec:lang:dataset}

To get a plausible ground truth
about the relative effects of skills and languages 
on code quality,
we analyze a subset of the data collected for a large-scale repository study~\citep{FSE}, as made available in \citet{TOPLAS}'s reanalysis.
This dataset summarizes the commit history of
hundreds of projects in various languages.

To keep things simple---and to avoid bias
that may come from underrepresented or misclassified languages~\citep{TOPLAS}---we only retain data about
projects:
\begin{enumerate*}
\item written in Python or Java (two widely used, yet fairly different languages);
\item with at least 100 commits on record;
\item that are not multi-language (that is, each project is entirely written in Java, or entirely written in Python).
\end{enumerate*}
These criteria select
105 projects: 45 written in Java and 60 written in Python.
For each of these projects, we are interested in two key variables:
\begin{description}
\item[\var{Language}:] a binary, ordinal variable (values: Java or Python),
  which denotes the project's language;
\item[\var{Quality}:] a continuous variable (ranging over $[0, 1]$),
  which measures the project's quality as the complement $1 - \var{Bugs}/\var{Commits}$ of the fraction of all project commits that
  are flagged as introducing a bug.\footnote{
    Naturally, this is a very crude, simplistic measure of quality.
    Again, this example's goal is not to discover new empirical knowledge
    about the fault proneness of different programming languages,
    but to illustrate how to apply technique to account for
    possible omitted variable bias.
  }
\end{description}

This dataset also includes information about
which developer produced each commit.
We use it to derive a crude estimate of the \emph{skills}
of developers active on a project as follows.
First, we only consider the 952 developers
who produced at least 10~commits each in the selected projects.
The skill of each developer $d$ among these ``frequent committers''
is the complement $1 - \var{Bugs}_d/\var{Commits}_d$
of the fraction of all commits authored by $d$ that introduced a bug.
Finally, variable \var{Skill} summarizes the skills
associated with each project:
\begin{description}
\item[\var{Skill}:] a continuous variable (ranging over $[0, 1]$),
  which is the mean skill of all developers among the ``frequent committers''
  who contributed to the project.
\end{description}

We stress that
the goal of this data selection process
is \emph{not} supporting any general claims
about the actual effects of a programming language on a project's quality.
It simply gives a rough idea
of the magnitude of these effects in a real world scenario,
so that we can appreciate that confounding is a plausible occurrence---hence, a practical concern.

\subsection{Quantifying Confounding}
\label{sec:lang:quantifying}

In this exercise,
the goal is estimating the effect of choosing
Java or Python as a programming language (variable \var{Language})
on the quality of the developed project (variable \var{Quality}).
If we had access also to variable \var{Skill},
we could single out the $\var{Language} \to \var{Quality}$ effect
by fitting \autoref{fig:m-qls}'s regression model $m_3$,
which uses \var{Language} and \var{Skill} as predictors.
On the data described in \autoref{sec:lang:dataset},
this produces an estimate of the effect $\var{Language} \to \var{Quality}$
of $\ell = -0.012$,
which indicates that using Python is very weakly
associated with a modest reduction of quality
(with a lot of uncertainty, as shown in~\autoref{tab:biased-unbiased}).

\begin{figure}[!tb]
  \centering
\begin{subfigure}{0.35\linewidth}
    \centering
    \begin{equation*}
      \begin{aligned}
        \var{Quality}_i &\sim \dist{Normal}(\mu_i, \sigma) \\
        \mu_i &= \alpha + \ell \cdot \var{Language}_i + s \cdot \var{Skill}_i
      \end{aligned}
    \end{equation*}
    \caption{Model $m_3$ with two predictors \var{Language} and \var{Skill}.}
    \label{fig:m-qls}
  \end{subfigure}
  \hfill
  \begin{subfigure}{0.3\linewidth}
    \centering
    \begin{equation*}
      \begin{aligned}
        \var{Quality}_i &\sim \dist{Normal}(\mu_i, \sigma) \\
        \mu_i &= \alpha + \ell \cdot \var{Language}_i
      \end{aligned}
    \end{equation*}
    \caption{Model $m_4$ with one predictor \var{Language}.}
    \label{fig:m-ql}
  \end{subfigure}
  \hfill
  \begin{subfigure}{0.3\linewidth}
    \centering
    \begin{equation*}
      \begin{aligned}
        \var{Quality}_i &\sim \dist{Normal}(\mu_i, \sigma) \\
        \mu_i &= \alpha + s \cdot \var{Skill}_i
      \end{aligned}
    \end{equation*}
    \caption{Model $m_5$ with one predictor \var{Skill}.}
    \label{fig:m-qs}
  \end{subfigure}
\\[2mm]
\begin{subtable}{0.65\linewidth}
  \begin{tabularx}{\linewidth}{cc r rr rr}
    \toprule
    \multicolumn{2}{c}{\textsc{model}}
    & \multicolumn{1}{c}{\textsc{estimate}}
    & \multicolumn{2}{c}{\textsc{95\% prob.}}
    & \multicolumn{2}{c}{\textsc{60\% prob.}}
    \\
    \midrule
    $m_3$ & uncounfounded & -0.012 & -0.035 & 0.011 & -0.021 & -0.002
    \\
    $m_4$ & confounded & -0.052 & -0.094 & -0.010 & -0.070 & -0.034
    \\
    \bottomrule
  \end{tabularx}
  \caption{Estimates of regression coefficient $\ell$ in models $m_3$ and $m_4$,
  with 95\% and 60\% probability intervals.}
  \label{tab:biased-unbiased}
\end{subtable}
  \caption{Different models to estimate the dependence between
    \var{Quality} and \var{Language}.}
  \label{fig:lang-quality-regressions}
\end{figure}

In practice, it may not be possible to reliably
measure the confounder \var{Skill}.
In this case,
we can only fit \autoref{fig:m-ql}'s model $m_4$,
which gives us a different estimate $\ell = -0.052$
of the effect
$\var{Language} \to \var{Quality}$
of \var{Language} on \var{Quality}.
If we compare this with the previous estimate based
on the unbiased model $m_3$,
we notice that the confounding of \var{Skill} \emph{inflates}
the effect $\var{Language} \to \var{Quality}$ as measured in this data.
In reality, we would not have access to the unbiased estimate;
how to assess how much confidence we can put in an
estimate that comes from a possibly confounded model?
There are three main ways of proceeding~\citep{linPK98sens}:
\begin{itemize}
\item If we can muster an
  estimate the confounding effect $\var{Skill} \to \var{Quality}$,
  we can use it to compute how strong the effect $\var{Skill} \to \var{Language}$
  would have to be to \emph{tip}
  the estimate of the $\var{Language} \to \var{Quality}$ effect
  (i.e., flip it from negative to positive).
  \autoref{sec:tip-SMD} discusses this scenario.

\item Conversely, if we can 
  estimate the confounding effect $\var{Skill} \to \var{Language}$,
  we can compute how strong the effect $\var{Skill} \to \var{Quality}$
  would have to be to tip the estimate
  of the $\var{Language} \to \var{Quality}$ effect.
  \autoref{sec:tip-effect} discusses this scenario.

\item Finally, assuming that there are several different confounders
  that simultaneously affect \var{Language} and \var{Quality},
  we can compute how many confounders with a certain effect would
  it take to tip the estimate
  of the $\var{Language} \to \var{Quality}$ effect.
  \autoref{sec:tip-num} discusses this scenario.
\end{itemize}

Assessing when tipping may occur under plausible scenarios
can inform us about the practical impact
of the confounder on our data analysis:
if tipping requires a strong confounding effect,
which is unlikely in practice,
we can probably tolerate the bias and still use our estimate, albeit imperfect,
of the effect of \var{Language} on \var{Quality}.
Conversely, if even a weak confounder is likely to tip,
we would conclude that we should not rely on the observational data analysis,
and try to collect different kinds of data
(or to use the existing data to answer different kinds of questions).
Such an analysis is useful also in more open-ended scenarios,
for example when we don't really know what (other) factors may affect
\var{Language} and \var{Quality}
but we have reasons to believe that \emph{some} unmeasured variables exist.
Conversely, tipping analyses are also applicable
when testing a hypothesis:
in such cases, the tipping value could indicate
the boundary between ``significant'' and ``not significant'' effect.

To perform these so-called tipping point sensitivity analyses
we will use the R package \texttt{tipr}~\citep{tipr},
which implements
state-of-the-art analysis techniques~\citep{linPK98sens}
based on
causal models
similar to those outlined in \autoref{sec:dags}.

\subsubsection{Scaled-mean Difference}
\label{sec:SMD}

In a tipping-point sensitivity analysis~\citep{tipr},
any unmeasured confounder $Z$ is modeled
as a standardized, normally distributed random variable.
For a binary treatment variable $X$ with two nominal values $X_0, X_1$,
this means that the impact of the confounder $Z$ on the treatment $X$
is captured by normal distributions with unit variance
and mean $\mu_0$ (when $X = X_0$) and $\mu_1$ (when $X = X_1$).

Accordingly, we use the \emph{scaled-mean difference} (\SMD)
to quantifying the impact of a confounder
(\var{Skill} in our example)
on the treatment
(\var{Language} in our example)
in a sensitivity analysis~\citep{tipr}.\footnote{
  The scaled-mean difference $\mu_1/\sigma_1 - \mu_2/\sigma_2$ between two groups
  is similar to, but different from,
  the standardized mean difference $(\mu_1 - \mu_2)/\sigma$---a commonly used
  effect size.
  Precisely, the scaled-mean difference is simply the difference of
  standardized means in the treatment and control groups.
}
In our case, the \smd{Skill}{Language}
of a programmer's skills in different language
is the difference between the mean \var{Skill} of a Python programmer
and the mean \var{Skill} of a Java programmer,
expressed as a multiple of \var{Skill}'s standard deviation:

\begin{equation}
  \label{eq:SMD}
  \smd{Skill}{Language} \quad=\quad
  \frac
  {\ev{\var{Skill} \mid \text{Python}}}{\sd{\var{Skill} \mid \text{Python}}}
    -
  \frac
  {\ev{\var{Skill} \mid \text{Java}}}{\sd{\var{Skill} \mid \text{Java}}}
\end{equation}
In \eqref{eq:SMD},
$\var{Skill} \mid x$
denotes
the values of \var{Skill} in all
datapoints where $\var{Language} = x$,
whereas
$\ev{D}$ denotes the mean and
$\sd{D}$ the standard deviation of some data $D$.

Since it's based on a standardized scale, the \SMD is easy
to interpret uniformly, across different settings and domains.
For instance, an \SMD of $\pm 3$ denotes a huge effect:
three standard deviations of difference
indicate that the distribution of skills of Python and Java programmers
barely overlap;
clearly, such a massive difference in skills is unlikely to
happen in practice.
In our dataset, $\smd{Skill}{Language}$ is $-1.545$,
which denotes quite a sizeable, albeit not huge, effect size.

\begin{table}[!bt]
  \centering
  \begin{tabular}{cr cr cr}
    \toprule
    \multicolumn{2}{c}{\textsc{measured}}
    & \multicolumn{2}{c}{\textsc{estimate}}
    & \multicolumn{2}{c}{\textsc{tipping}}
    \\
    \cmidrule(lr){1-2}
    \cmidrule(lr){3-4}
    \cmidrule(lr){5-6}
    effect $\var{Language} \to \var{Quality}$ & -0.052
    & effect $\var{Skill} \to \var{Quality}$ & 0.835
    & \SMD $\var{Skill} \to \var{Language}$ & -0.062
    \\\midrule
    effect $\var{Language} \to \var{Quality}$ & -0.052
    & \SMD $\var{Skill} \to \var{Language}$ & -1.545
    & effect $\var{Skill} \to \var{Quality}$ & 0.034
    \\\midrule
    \multirow{2}{*}{effect $\var{Language} \to \var{Quality}$}
    & \multirow{2}{*}{-0.052}
    & effect $\var{C} \to \var{Quality}$ & 0.170
    & \multirow{2}{*}{number of confounders \var{C}}
    & \multirow{2}{*}{2}
    \\
    & & \SMD $\var{C} \to \var{Language}$ & -0.150
    \\
    \bottomrule
  \end{tabular}
  \caption{Three scenarios where we calculate the effect sufficient to produce \textsc{tipping} the \textsc{measured} effect $\var{Language} \to \var{Quality}$,
    based on an \textsc{estimate} of the confounder \var{Skill} on
    either the outcome (first row) or the treatment (second row),
    or of several unknown confounders \var{C} (bottom rows).
  }
  \label{tab:tipping-summary}
\end{table}

\subsubsection{Confounding \SMD}
\label{sec:tip-SMD}

In this scenario, we have measured the (possibly confounded)
effect $\var{Language} \to \var{Quality}$,
and we have an idea (for instance, from other studies)
of a plausible value for the
confounding effect $\var{Skill} \to \var{Quality}$.
From this data,
we calculate the \smd{Skill}{Language} that would
lead to a confounding such that our estimate of the effect
$\var{Language} \to \var{Quality}$ has the opposite sign of the ``true'' effect.

In our running example,
the estimate of $\ell = -0.052$ using model $m_4$
is the measured effect $\var{Language} \to \var{Quality}$;
whereas the estimate of $s = 0.835$ using \autoref{fig:m-qs}'s model $m_5$
gives us a plausible value for the confounding effect $\var{Skill} \to \var{Quality}$.
As shown in \autoref{tab:tipping-summary},
a modest \smd{Skill}{Language}
of $-0.062$
would be sufficient to flip the sign of the measured effect.
Such an \SMD is fairly modest, and likely to happen in practice;
in fact, we have seen that the \SMD measured in the data is much larger.
In all, we cannot have much confidence that the estimate
of the effect $\var{Language} \to \var{Quality}$
is valid.

\subsubsection{Confounding Effect}
\label{sec:tip-effect}

In this scenario, we have measured the (possibly confounded)
effect $\var{Language} \to \var{Quality}$,
and we have an idea (for instance, from other studies)
of a plausible value for the
confounding \SMD $\var{Skill} \to \var{Language}$.
From this data,
we calculate the effect $\var{Skill} \to \var{Quality}$ that would
lead to a confounding such that our estimate of the effect
$\var{Language} \to \var{Quality}$ has the opposite sign of the ``true'' effect.

In our running example,
the estimate of $\ell = -0.052$ using model $m_4$
is, once again, the measured effect $\var{Language} \to \var{Quality}$;
whereas the $\smd{Skill}{Language} = -1.545$
measured on the data according to \eqref{eq:SMD}
gives us a plausible value
for the confounding effect $\var{Skill} \to \var{Language}$.
As shown in \autoref{tab:tipping-summary},
a modest effect $\var{Skill} \to \var{Quality}$
of $0.034$
would be sufficient to flip the sign of the measured effect.
Such a confounding effect is fairly modest, and likely to happen in practice;
in fact, we have seen that the estimate
of this effect $s = 0.835$ using \autoref{fig:m-qs}'s model $m_5$ is much larger.
Also in this scenario, we cannot have much confidence that the estimate
of the effect $\var{Language} \to \var{Quality}$
is valid.

\subsubsection{Number of Confounders}
\label{sec:tip-num}

In this scenario, we have measured the (possibly confounded)
effect $\var{Language} \to \var{Quality}$,
and we are considering several different confounders.
We have an idea
of a plausible value for both the
\SMD $\var{C} \to \var{Language}$
and the effect $\var{C} \to \var{Quality}$
for any such unknown confounders \var{C}.
From this data,
we calculate how many such variables \var{C}
would produce an overall confounding such that our estimate of the effect
$\var{Language} \to \var{Quality}$ has the opposite sign of the ``true'' effect.

As usual,
the estimate of $\ell = -0.052$ using model $m_4$
serves as the measured effect $\var{Language} \to \var{Quality}$.
Then,
we can speculate that
a generic confounder \var{C}
has an $\smd{C}{Language} = -0.15$
and an effect $\var{C} \to \var{Quality} = 0.17$.
These values are respectively $1/10$ and $1/5$
of the corresponding values for \var{Skill}
as measured in the dataset;
intuitively, they represent confounders with a much more tamed power
compared to \var{Skill}.
Nevertheless,
just two such generic confounders
would be sufficient to flip the sign of the measured effect.
Since it is definitely plausible that there are
a couple of confounders with moderate effect,
we reach again the same conclusion
that we cannot have much confidence that the estimate
of the effect $\var{Language} \to \var{Quality}$
is valid.

\begin{figure}[tb]
  \centering
  
    \includegraphics[width=0.85\linewidth]{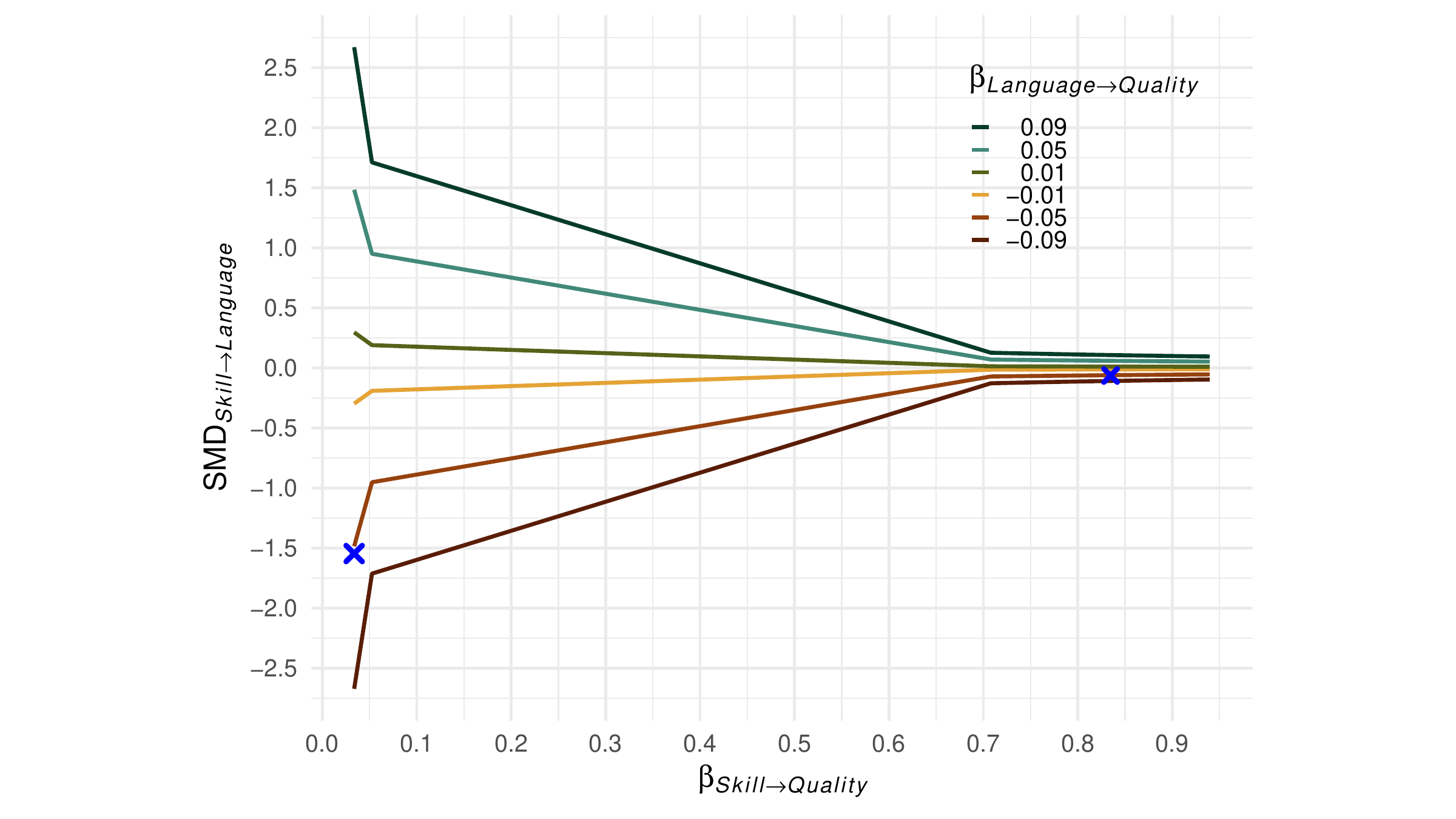}
    \caption{A graphical summary of tipping analyses of confounder effect of \var{Skill} on the $\var{Language} \to \var{Quality}$ effect.
      For different measured values of effect ($\beta_{\var{Language} \to \var{Quality}}$),
    the plot shows the values of the effect $\var{Skill} \to \var{Quality}$
    ($\beta_{\var{Skill} \to \var{Quality}}$)
    and of the \SMD $\var{Skill} \to \var{Language}$ (labeled $\textsf{SMD}_{\var{Skill} \to \var{Language}}$ in the plot)
    that would tip the main effect.
    The two blue cross marks correspond to the two scenarios discussed
    in \autoref{sec:tip-SMD} and \autoref{sec:tip-effect}.
  }
    \label{fig:tipping-lang-quality}
\end{figure}

\subsection{Sensitivity Analysis}
\label{sec:lang:sensitivity}

Let's generalize the tipping analysis described in this section
beyond \citet{FSE}'s data.
First of all,
consider a range of possible measured effects
$\var{Language} \to \var{Quality}$,
including both positive and negative values.
For each of them,
\autoref{fig:tipping-lang-quality}
plots a line of a different color
that marks the combination of values
of confounding effect $\var{Skill} \to \var{Quality}$
(horizontal axis)
and \SMD $\var{Skill} \to \var{Language}$
(vertical axis)
that would tip the measured effect.

\autoref{fig:tipping-lang-quality} indicates that
the two confounders,
$\beta_{\var{Skill} \to \var{Quality}}$
and
$\SMD_{\var{Skill} \to \var{Language}}$
are inversely proportional.
Intuitively, this is because
both confounding factors
concur to introduce tipping, but each affects a different variable;
thus, 
the more pronounced one of them is,
the less it is required of the other.
Another observation is that the various colored lines in \autoref{fig:tipping-lang-quality}
become flatter as the absolute value of the measured effect
$\beta_{\var{Language} \to \var{Quality}}$ decreases.
In fact,
if the measured effect is small,
even moderate-magnitude confounders may introduce a strong bias;
whereas stronger effects are swayed only by
correspondingly strong confounders.

Such a sensitivity analysis can provide a useful
guide not only to analyze empirical data,
but to \emph{plan} new experiments to improve
the validity of existing findings.
For example,
it would be interesting to collect reliable data
about the relations $\var{Skill} \to \var{Quality}$
(what's the impact of a developer's skills on the quality of code they produce?)
and $\var{Skill} \to \var{Language}$
(do developers with different skill have marked preferences for which language
to use?).
Furthermore,
a sensitivity analysis
would enhance the value of an empirical study for other researchers,
since it would better, and quantitatively,
identify the study's envelope of validity,
and it would make the study's assumptions more transparent.

\section{Confounding in Teamwork Effort}
\label{sec:teamwork}
\label{sec:2ndcase}

Our second illustrative example is based on
\citet{feldtSFT23sysrev}'s
review of research in the factors
that affect the productivity of software development teams.
It demonstrates how to
practically assess the impact of potential unknown confounders
in a more complex, realistic scenario.
As in \autoref{sec:lang:quantifying}'s illustrative example,
our aim is not to (directly) contribute
to the knowledge about developer productivity;
in fact, our analysis will at times adopt simplifying assumptions 
that are not (entirely) realistic.
In contrast,
the goal is to demonstrate credible confounding patterns,
and to illustrate how to navigate around them in
realistic conditions.

\begin{figure}[!b]
  \centering
  \begin{tikzpicture}[every node/.style={minimum size=7mm}]
    \matrix[row sep=7mm,column sep=20mm] {
      & \node[plain,confounder] (n-S) {\var{S}}; & & & \node[plain,treatment] (n-T) {\var{T}};
      \\
      & \node[plain] (n-K) {\var{K}}; & & \node[plain] (n-D) {\var{D}};
      \\
      \node[plain] (n-B) {\var{B}}; & & & \node[plain,outcome] (n-E) {\var{E}};
      \\
      & \node[plain] (n-H) {\var{H}}; & & \node[plain] (n-L) {\var{L}};
      \\
      & \node[plain] (n-P) {\var{P}}; & & & \node[plain,confounder] (n-O) {\var{O}};
      \\
    };

    \foreach \nfrom/\nto in {
      O/P,
      B/P,
      H/P,
      P/E,
      O/L,
      L/E,
      O/E,
      B/H,
      H/E,
      B/E,
      O/T,
      B/S,
      K/S,
      K/E,
      D/E,
      S/E,
      S/T,
      T/E} {
      \draw[->, thick] (n-\nfrom) -- (n-\nto);
  }

\end{tikzpicture}
\caption{A DAG summarizing relations among variables that characterize a software project, based on \citet{feldtSFT23sysrev}'s literature review.
    The variables are, from top to bottom and left to right:
    software \var{S}ize,
    \var{T}eam size,
    software \var{K}ind,
    \var{D}omain,
    \var{B}usiness area,
    \var{E}ffort,
    \var{H}ardware,
    \var{L}ocation,
    \var{P}rogramming language,
    and \var{O}rganization type.
  }
  \label{fig:UDAG-1}
  \label{fig:productivity-dag}
\end{figure}

\subsection{Data: Teamwork, Effort, and Other Covariates}
\label{sec:productivity-dag}

\citet{feldtSFT23sysrev}
propose to use causal DAGs to summarize and combine
the key findings of systematic literature reviews.
In one of their case studies,
they review several primary studies
about productivity in software development.
The DAG shown in \autoref{fig:productivity-dag}
is one of the DAGs that is obtained
by applying \citet{feldtSFT23sysrev}'s
approach;
it summarizes the main relevant relations between variables
that have been observed in some of the reviewed literature on the topic.\footnote{
  This does not mean that this is the ultimate summary of research in the area
  of software development productivity. For our purposes,
  all that matters is that it displays a rich collection of \emph{plausible}
  relations, so that our omitted variable bias analysis
  is grounded in a realistic scenario.
  }

  \autoref{fig:productivity-dag}'s DAG
  consists of 18 relations (arrows) among 10 variables:
\begin{description}
\item[\var{B}:] the company's \textbf{b}usiness area
  (accounting, sales, human resources, \ldots)
\item[\var{D}:] the project's \textbf{d}omain
  (healthcare, finance, entertainment, \ldots)
\item[\var{E}:] the overall \textbf{e}ffort spent by the software developers
  (hours worked, planned years, \ldots)
\item[\var{H}:] the \textbf{h}ardware that runs the software
  (server, client, mobile, embedded, \ldots)
\item[\var{K}:] the \textbf{k}ind of software
  (application, library, system, web, \ldots)
\item[\var{L}:] the company's \textbf{l}ocation
  (North America, Europe, Asia, \ldots)
\item[\var{O}:] the company's \textbf{o}rganization type
  (private, public, non-profit, \ldots)
\item[\var{P}:] the project's \textbf{p}rogramming language
\item[\var{S}:] the software \textbf{s}ize
  (lines of code, function points, \ldots)
\item[\var{T}:] the size of the \textbf{t}eam of programmers on the project
\end{description}

In the rest of this section, we put ourselves in the shoes
of a researcher who is designing a new study
to determine the strength of the causal
relation between \var{T} (team size) and \var{E} (effort).
The key questions that need to be addressed
to design such a study are:
\begin{enumerate}
\item \label{q:adj-analysis}
  What variables, other than the treatment \var{T} and the outcome \var{E},
  ought to be measured?

\item \label{q:conf-analysis}
  Given the variables that could be effectively measured,
  what possible remaining confounders
  of the causal effect of treatment \var{T} on outcome \var{E} may remain?
\end{enumerate}

Since \autoref{fig:productivity-dag}
summarizes several primary studies in the domain of
software development productivity,
we can conveniently use it as the basis of further studies in the same domain.
Even if we were targeting a domain with little
prior research,
we could still build a DAG that captures
whatever is known based on the state of the art in this domain.
A DAG is just a convenient notation to
summarize knowledge about the structural relations among variables;
if previous work is scarce,
the DAG will be simplistic or incomplete
but will still serve as a useful guide.
At a minimum,
we can always fall back to
building a minimal DAG such as in \autoref{fig:XYZX},
which just captures
the relation of interest $X \to Y$ and a generic confounder $Z$.

\subsection{Adjustment Sets}
\label{sec:adjustment-sets}

As in the previous illustrative example,
to estimate the effect of a treatment
(\var{T} in our example)
on an outcome
(\var{E} in our example),
we fit on the data a linear regression model $m_A$:
\begin{equation}
  \begin{aligned}
    \var{E}_i &\sim \dist{Normal}(\mu_i, \sigma) \\
    \mu_i &= \alpha + \beta_t \cdot \var{T} + \sum_{v \in A} \beta_v \cdot v_i
  \end{aligned}
  \label{eq:mA}
\end{equation}
The estimate of coefficient $\beta_t$ in the fitted model $m_A$ measures the effect $\var{T} \to \var{E}$.

Thus, addressing question \autoref{q:adj-analysis} above
(\emph{What variables ought to be measured?})
is tantamount to deciding which variables (covariates)\footnote{
  In common statistical jargon, a covariate is
  any predictor variable other than the treatment.
  }
should be included in set $A$ in model $m_A$ \eqref{eq:mA}.
In particular, $A$ should include all \emph{confounders},
so that the effect $\var{T} \to \var{E}$ can be estimated without bias.
Set $A$ represents
what is called an \emph{adjustment set}:
given a DAG and two of its variables---representing the outcome (\var{E} in our case)
and the treatment (\var{T} in our case)---the adjustment set $A$ is the set of additional variables in the DAG
that we have to include in \eqref{eq:mA}'s model
to ensure that $\beta$ estimates the unbiased, uncounfounded,
genuine causal effect of \var{T} on \var{E}.
In other words, the adjustment set
includes all predictors that we should include
if we want to avoid introducing omitted variable bias.

The adjustment set
(or rather adjustment sets,
since a DAG may admit multiple, alternative adjustment sets)
can be computed directly on the DAG based solely on its structure---assuming, of course, that the DAG correctly captures
real-world causal relations.\footnote{
  In previous work~\citep{furiaTF23causal},
  we illustrated how adjustment sets are computed,
  and what's the intuition behind them.
  For brevity, we do not repeat the explanation in this paper,
  but simply \emph{use} the adjustment sets computed from a DAG
  in our analysis.
}
In our example,
the adjustment set is $A = \{ \var{O}, \var{S} \}$;
therefore including the two predictors \var{O} and \var{S},
in addition to \var{T}, 
ensures that $\beta$ captures the net causal relation between
treatment and outcome.

Remember that the adjustment set is a variable selection strategy
whose goal is correcting for possible confounding effects
of causal relations.
As explained in \autoref{sec:scope-limitations},
this is the specific focus of this paper;
naturally, 
different kinds of research questions
may require different variable selection criteria.\footnote{
  In particular, if the goal is fitting a statistical (regression) model
  with an optimal trade off between predictive accuracy and model conciseness,
  information-theoretic model selection criteria provide a serviceable approach~\citep{Bayesian-varselect,watanabe10waic,vehtariGG17loo}.
}
This does not necessarily imply that adjusting for confounding bias
is irreconcilable with other data analysis goals.
In fact, we will see that a DAG often admits different equivalent
adjustment sets,
which often provides more flexibility in how this technique is applied.

\subsection{Unmeasured Confounders}
\label{sec:unmeasured-confounders}

Even though \autoref{fig:productivity-dag}'s DAG
is based on several empirical studies,
it still is completely plausible
that it does not include all factors that contribute
to the observed relation between treatment and outcome.
In fact, in every complex, real-world process,
it is exceedingly likely that there are unmeasured
variables that might still have a sizeable impact
on the variables of interest.

\begin{table}[!tb]
  \centering
  \begin{tabular}{rc r}
    \toprule
    \multicolumn{2}{c}{\textsc{confounded edge}} & \textsc{adjustment sets} \\
    \midrule
    1 & $B \to E$ & $\{ O, S\}$  \\
    2 & $B \to H$ & $\{ O, S\}$  \\
    3 & $B \to P$ & $\{ O, S\}$  \\
    4 & $B \to S$ & $\{ O, S\}$  \\
    5 & $D \to E$ & $\{ O, S\}$  \\
    6 & $H \to E$ & $\{ O, S\}$  \\
    7 & $H \to P$ & $\{ O, S\}$  \\
    8 & $K \to E$ & $\{ O, S\}$  \\
    9 & $K \to S$ & $\{ O, S\}$  \\
    10 & $L \to E$ & $\{ O, S\}$  \\
    11 & $O \to E$ & $\{ O, S\}$  \\
    12 & $O \to L$ & $\{ O, S\}$  \\
    13 & $O \to P$ & $\{ O, S\}$  \\
    14 & $O \to T$ & $\{ O, S\}$  \\
    15 & $P \to E$ & $\{ O, S\}$  \\
    16 & $S \to E$ & $\{ O, S\}$  \\
    17 & $S \to T$ & $\{ B, K, O, S\}$ $\{ O, S, Z_{S, T}\}$  \\
    18 & $T \to E$ & $\{ O, S, Z_{T, E}\}$     \\
    \bottomrule
  \end{tabular}
  \caption{Adjustment sets for \autoref{fig:productivity-dag}'s DAG extended with a confounder $Z_{X, Y}$ affecting each edge $X \to Y$.}
  \label{tab:adjustment-sets}
\end{table}

To address this issue of unmeasured (unknown) additional confounders,
we can extend the previous adjustment set analysis.
For every pair of nodes $X \to Y$ in \autoref{fig:productivity-dag}'s DAG,
we introduce an unmeasured confounder $X \leftarrow Z_{X, Y} \to Y$
that affects $X$ and $Y$ simultaneously.
Then, we recompute the adjustment set of the DAG extended with such
additional node.
\autoref{tab:adjustment-sets} shows the results of this analysis.

Adding a confounder to 16 out of 18 edges in \autoref{fig:productivity-dag}'s DAG
does not change the adjustment set, which remains $\{ O, S \}$---as when considering the original DAG without unknown confounders.
In other words, including $O$ and $S$ in model $m_A$ \eqref{eq:mA}
conveniently also voids the effect of other possible confounders.
In contrast,
a confounder $Z_{S, T}$ affecting edge $S \to T$ admits two adjustment sets,
one that includes $Z_{S,T}$ itself, and one that does not.
Clearly, we prefer the latter adjustment set $\{ B, K, O, S \}$:
since it does not include unknown variable $Z_{S,T}$,
it ensures that any possible confounding effect of $Z_{S,T}$
is corrected for indirectly, even if $Z_{S,T}$ cannot be measured---in fact, without even knowing what this variable represents.

Unfortunately, the last edge $T \to E$
cannot be handled so easily.
If there were an additional variable $Z_{T,E}$ simultaneously
affecting $T$ and $E$,
the only way to correct its confounding effect
would be to measure $Z_{T,E}$ and include it as a predictor in \eqref{eq:mA}'s model.
By definition, this is impossible because
we don't know what $Z_{T,E}$ is, or we have an idea but
it is impractical or impossible to measure it.

\begin{figure}[!tb]
  \centering
  \begin{tikzpicture}[every node/.style={minimum size=7mm}]
    \matrix[row sep=7mm,column sep=20mm] {
      & \node[confounder] (n-S) {\var{S}}; & & & \node[treatment] (n-T) {\var{T}};
      \\
      & \node[confounder] (n-K) {\var{K}}; & & \node[unmeasured] (n-Z) {\var{Z}};
      \\
      \node[confounder] (n-B) {\var{B}}; & & & \node[outcome] (n-E) {\var{E}};
      \\
      &  & & ;
      \\
      &  & & & \node[plain,confounder] (n-O) {\var{O}};
      \\
    };

    \foreach \nfrom/\nto/\lab in {
      Z/E/$\gamma_e$,
      Z/T/$\gamma_t$,
      O/E/$o_e$,
      B/E/$b_e$,
      O/T/$o_t$,
      B/S/$b_s$,
      K/S/$k_s$,
      K/E/$k_e$,
      S/E/$s_e$,
      S/T/$s_t$,
      T/E/$t_e$} {
      \draw[->, thick] (n-\nfrom)
      -- node[fill=white,inner sep=0pt, minimum size=0pt]
       {\lab} (n-\nto);
    }
\end{tikzpicture}
\caption{~\autoref{fig:productivity-dag}'s DAG simplified to include only
  treatment $T$, outcome $E$, as well as all variables in the adjustment set
  that also accounts for unknown/unmeasured confounders on other edges.
  The DAG also shows an unmeasured confounder $Z$ that may still exist:
  we cannot adjust it away with any other variables.}
    \label{fig:UDAG-2}
    \label{fig:UDAG-adjusted}
\end{figure}

\subsection{Sensitivity Analysis}
\label{sec:sensitivity-E-T}

\autoref{fig:UDAG-adjusted}
summarizes the analysis results so far:
in order to get an unbiased estimate of the $T \to E$
causal relation in \autoref{fig:productivity-dag}'s DAG,
we should include $B, K, O, S$ as additional predictors,
which safeguards against unmeasured confounders
affecting other relations in the DAG.
However, there remains a possible confounder $Z_{T, E}$---which we'll just call $Z$ from now on---that cannot be controlled for indirectly by means of other known variables.

Whether $Z$'s confounding is negligible or consequential
ultimately depends on its strength relative to the
strength of the other causal relations.
Roughly, if $E \to T$ and the other relations are very strong,
whereas $Z \to E$ and $Z \to T$ are very weak,
our estimate of the $E \to T$ relation has a good chance
of remaining reliable
even if we cannot measure the unknown $Z$.
Our next analysis step is thus a sensitivity analysis
based on \emph{simulation}:
using some plausible, informed estimates for the
various effects in the DAG,
we'll try to recover the effect $E \to T$ without
measuring $Z$, and we'll see how far off this estimate is from the
ground truth.

\subsubsection{Simulation Parameters}

For our simulation,
we use the following generative model,
which mirrors the structure of \autoref{fig:UDAG-adjusted}'s DAG:
\begin{equation}
  \begin{aligned}
    B       &\sim   \dist{Binomial}(1, 0.5) &
    K       &\sim   \dist{Binomial}(1, 0.5) \\
    O       &\sim   \dist{Binomial}(1, 0.5) &
    Z       &\sim   \dist{Normal}(0, 1) \\
S       &\sim   \dist{Normal}(b_s B + k_s K, 1) &
    T       &\sim   \dist{Normal}(o_t O + s_t S + \gamma_t Z, 1) \\
    E       &\sim   \dist{Normal}(b_e B + k_e K + o_e O + s_e S + t_e T + \gamma_e Z, 1)
  \end{aligned}
  \label{eq:gen-model}
\end{equation}
In \eqref{eq:gen-model},
every categorical variable is binary
and follows a Bernoulli distribution
with $0.5$ probability of drawing a $1$;
the other variables are real-valued and follow
a normal distribution with unit variance
and mean that is given by a linear combination of the variables
that directly affect it according to \autoref{fig:UDAG-adjusted}'s model.
This is admittedly a strongly simplified generative model,
but it has the advantage that we can choose
standardized effect sizes for its parameters,
instead of having to rely on difficult-to-obtain
estimates on a natural scale.
Furthermore, \eqref{eq:gen-model}
could be generalized (e.g., to include
categorical variables with more than two possible values),
or specialized according to domain-specific characteristics
(e.g., to use truncated distributions that capture hard bounds of the value of some variables).

The generative model in \eqref{eq:gen-model}
has 11 parameters, one for each edge in \autoref{fig:UDAG-adjusted}'s DAG.
Each parameter $x_y$ denotes the effect of $X$ on $Y$, corresponding to
edge $X \to Y$.
The exceptions are edges $Z \to E$ and $Z \to T$, whose
effects we denote as $\gamma_e$ and $\gamma_t$ to single them out
(as they are the part of the model that is completely unmeasured).
In a concrete case study,
the simulation would use parameters
that reflect the system that
is actually being observed, where the data comes from.
In our case,
we do not have a specific data collection process in mind.
Alternatively, one could simply try out all parameter
combinations that are remotely plausible.
In our case,
such an exhaustive analysis would be
practically infeasible;
for example, if each parameter can take 6 possible values,
we would end up with $6^{11}$ parameter combinations---that is over 362 millions!

Instead,
we go back to \citet{feldtSFT23sysrev} and use the statistics from some of the reviewed primary studies
to get a ballpark estimate of the relevant effects.
This trades off some generality for a manageable simulation time.
\autoref{tab:sim-params}
shows the range of parameter values that we used for our simulations.
\begin{description}
\item[Confounder:] First of all, we want to simulate all plausible confounding
  scenarios of $Z$; therefore, we consider all combinations of small
  (0.1), medium (0.3), and large (0.5) standardized effect sizes\footnote{The values 0.1, 0.3, and 0.5 are often considered as the boundaries
    between
    negligible, small, medium, and large effect
    sizes
    for effects measured
    as standardized mean differences~\cite[p.~224--225]{Cohen-book}~\cite{effect-size-article}.
    This is just a plausible choice of values for our illustrative example;
    in concrete case studies, one could pick parameter values more precisely,
    based on what is considered a small, medium, or large effect
    in the domain of interest.
  }---both positive and negative---for each parameter $\gamma_e$ and $\gamma_t$.

\item[Main effect:] We also consider all
  possible effect sizes for parameter $t_e$, which represents the
  treatment effect that we are trying to estimate; however, we only
  consider \emph{positive} effect sizes since, according to the
  studies reviewed in \citet{feldtSFT23sysrev}, it is implausible that
  large teams produce an overall lesser effort than small teams.

\item[Indirect effects:] As for the effects $x_e$ corresponding to
  the edges $X \to E$, for every other variable $X$ in the adjustment set,
  we picked a small, medium, or large, positive or negative
  effect size based on some of the studies reviewed in \citet{feldtSFT23sysrev}.
  Since those studies focused on effort (or related variables)
  as outcome, we could not find any hard data
  about the magnitude of the effects of these other variables $X$
  on \emph{other} covariates.
  Simplistically, we assume that $x_y$ is the same as $x_e$,
  that is variable $X$ has roughly the same effect
  on all variables it directly affects.
  Again, a specific case study could come up with
  more definite estimates; our goal is mainly
  to demonstrate this analysis method on somewhat plausible data.
\end{description}

\begin{table}[!tb]
  \centering
  \begin{tabular}{lc rl}
    \toprule
    \multicolumn{2}{c}{\textsc{parameter}}
    & \multicolumn{1}{c}{\textsc{values}}
    & \multicolumn{1}{c}{\textsc{justification}}
    \\
    \midrule
    $b_e$ & $B \to E$ & 0.3
    & \cite[Tab.~V, $\omega^2$/maintenance (all)]{TsunodaOno}
    \\
    $b_s$ & $B \to S$ & 0.3
    & Same as $b_e$
    \\
    $k_e$ & $K \to E$ & 0.1
    & \cite[Tab.~XIII, \% variance explained/project type]{WangEtAl}
    \\
    $k_s$ & $K \to S$ & 0.1
    & Same as $k_e$
    \\
    $o_e$ & $O \to E$ & 0.5
    & \cite[Tab.~IV, $\omega^2$/maintenance (all)]{TsunodaOno}
    \\
    $o_t$ & $O \to T$ & 0.5
    & Same as $o_e$
    \\
    $s_e$ & $S \to E$ & -0.1
    & \cite[Tab.~XIII, $\rho$/maintenance (all)]{TsunodaOno}
    \\
    $s_t$ & $S \to T$ & -0.1
    & Same as $s_e$
    \\
    $t_e$ & $T \to E$ & 0.1, 0.3, 0.5
    & all plausible positive effect sizes
    \\
    $\gamma_e$ & $Z \to E$ & -0.5, -0.3, -0.1, 0.1, 0.3, 0.5
    & all plausible effect sizes
    \\
    $\gamma_t$ & $Z \to T$ & -0.5, -0.3, -0.1, 0.1, 0.3, 0.5
    & all plausible effect sizes
    \\
    \midrule
$n$ & sample size & 5, 10, 50
          &
    \\
    $n_{\textrm{sim}}$ & repetitions
    & 200
    \\
    \bottomrule
  \end{tabular}
  \caption{Range of values for the parameters of the generative model in \eqref{eq:gen-model} used in the simulation. For each parameter,
    the table also reports the \textsc{justification} for the choice of \textsc{values}, usually as a reference to a primary study that
  measured such an effect or a comparable one.}
  \label{tab:sim-params}
\end{table}

The simulation includes implicitly two more parameters:

\begin{description}
\item[Sample size:] the sample size $n$ determines how many
  datapoints we sample from \eqref{eq:gen-model}'s generative model.
  We try three different sample sizes: 5, 10, and 50.
  Due to the nature of the data we are simulating,
  small sample sizes (i.e., 5 and 10) are especially relevant
  and realistic: collecting
  all such detailed data about many software projects
  would be costly (in particular, it's unlikely
  that such data can be reliably obtained
  by simply mining open-source repositories);
  hence, an actual empirical study would likely be limited
  to a smallish sample size.
  Nevertheless, we also include a more substantial sample size
  (i.e., 50) to extend the reach of our analysis.

\item[Repetitions:] For each parameter combination,
  we repeat the whole simulation-inference process
  $n_{\textrm{sim}}$ times, and take the average of the obtained estimates.
  We go with 200 repetitions, which should be enough
  to smoothen out any random fluctuations in our simulations.
\end{description}

\subsubsection{Simulation Process}

For every combination of values for the parameters in \autoref{tab:sim-params},
\autoref{algo:sim} uses generative model \eqref{eq:gen-model}
to draw random samples of the
variables $B, K, O, Z, S, T, E$ (consistently with the dependencies shown in \autoref{fig:UDAG-adjusted}).
It then fits the following linear regression model
on the samples:
\begin{equation}
  \begin{aligned}
    \var{E}_i &\sim \dist{Normal}(\mu_i, \sigma) \\
    \mu_i &= \alpha + \beta_t \cdot \var{T}_i + \beta_b \cdot \var{B}_i + \beta_k \cdot \var{K}_i + \beta_o \cdot \var{O}_i + \beta_s \cdot \var{S}_i
  \end{aligned}
  \label{eq:inf-sim}
\end{equation}
Model \eqref{eq:inf-sim}
estimates the effect size $\beta_t$ of $T$ on $E$
without conditioning on $Z$, which we assume cannot be measured,
but otherwise includes all measurable variables in the adjustment set.

This sample\slash fit process is repeated $n_{\textrm{sim}}$ times
for each parameter combination;
finally, \autoref{algo:sim}
returns the mean $\beta_t$,
the 50\% highest-probability density interval $\ell_{50} \ldots u_{50}$,
and the 95\% highest-probability density interval $\ell_{95} \ldots u_{95}$
of the estimates of $\beta_t$ over all repetitions.\footnote{
  Highest density intervals are a Bayesian analog to confidence intervals
  in frequentist statistics~\cite{BDA3,conf-vs-bayes}.
  Since highest density intervals are simply defined as intervals on
  a probability distribution, their interpretation
  is usually considered
  more intuitive~\cite{robustMisinterpretation,Hespanhol2019Intervals}.
}
These statistics summarize the likely ranges of $\beta_t$ estimated from the simulations.

\begin{figure}[!htb]
  \centering
  \includegraphics[width=1.0\linewidth]{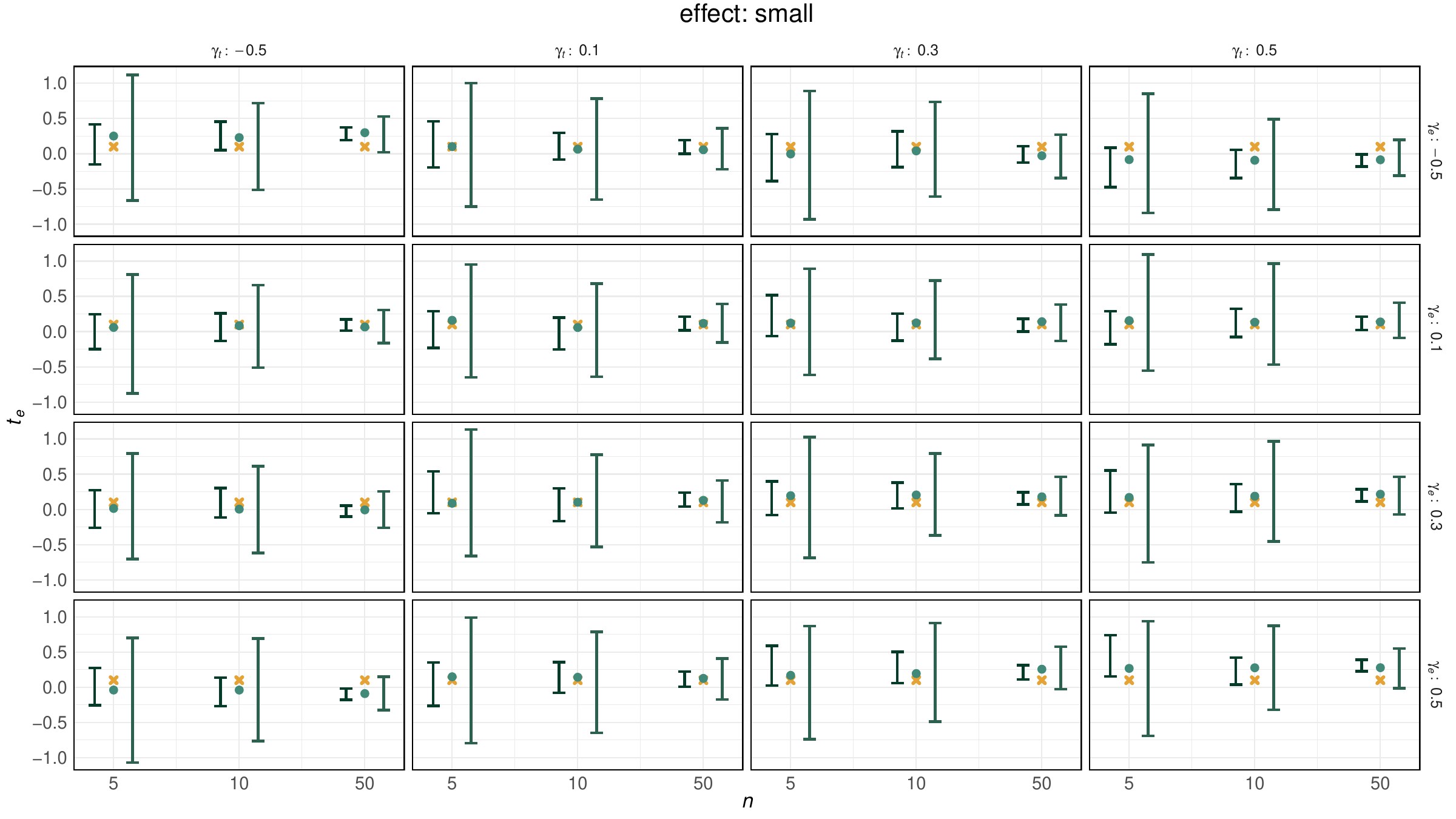}
  \caption{A summary of the results of running \autoref{algo:sim} for some parameter combinations in \autoref{tab:sim-params} and \emph{small} ground truth effect $t_e = 0.1$. The plots report, for each sample size 5, 10, 50,
    the ground truth $t_e$ \gtmark,
    the mean estimate \estmark,
    the 50\% probability interval \errorbar[degascolg] (shifted left),
    and the 95\% probability interval \errorbar[degascolf][0.9ex] (shifted right)
    of $\beta_t$.}
  \label{fig:sim-plots-small}
\end{figure}

\begin{figure}[!htb]
  \centering
  \includegraphics[width=1.0\linewidth]{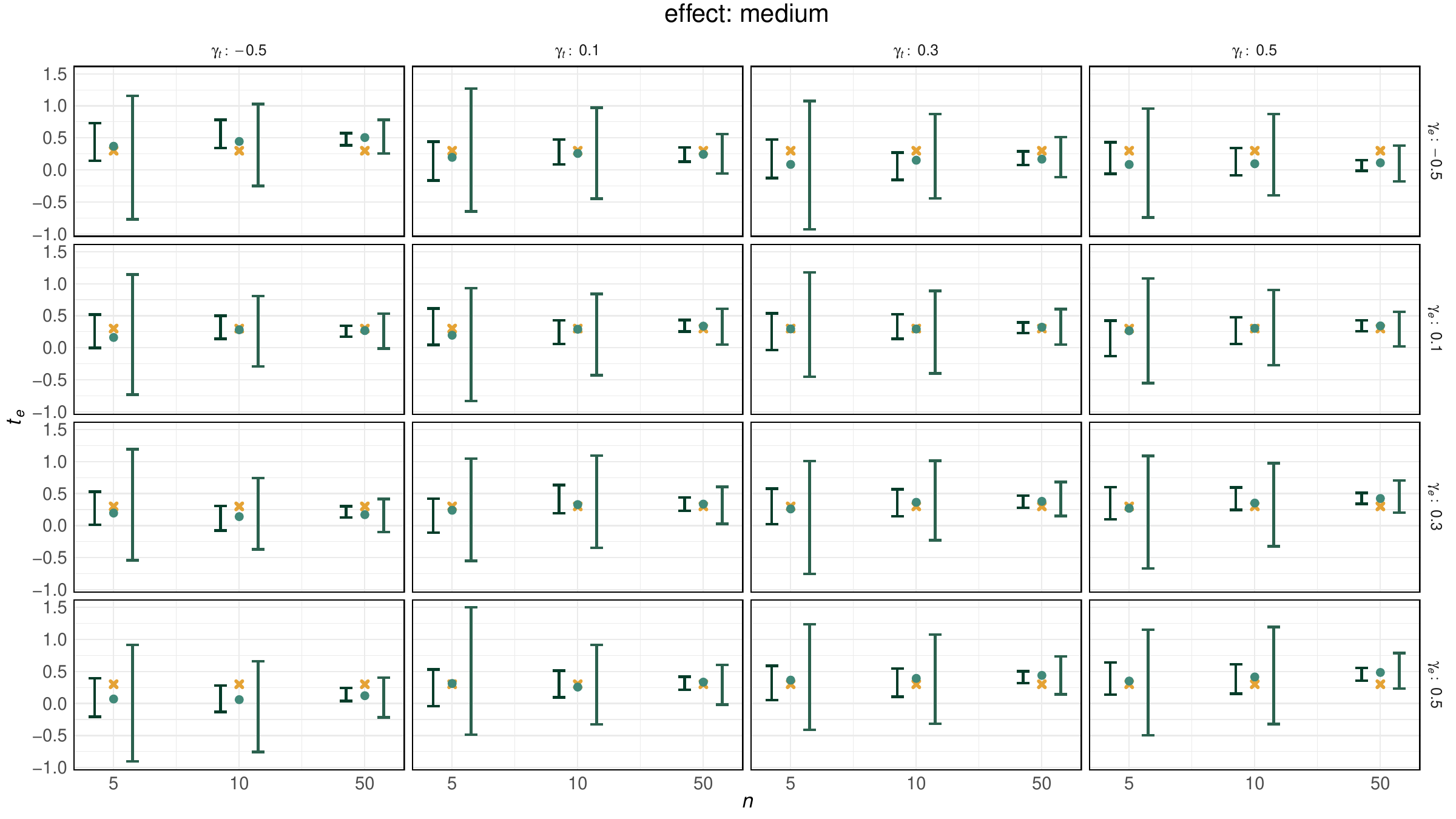}
  \caption{A summary of the results of running \autoref{algo:sim} for some parameter combinations in \autoref{tab:sim-params} and \emph{medium} ground truth effect $t_e = 0.3$. The plots report, for each sample size 5, 10, 50,
    the ground truth $t_e$ \gtmark,
    the mean estimate \estmark,
    the 50\% probability interval \errorbar[degascolg] (shifted left),
    and the 95\% probability interval \errorbar[degascolf][0.9ex] (shifted right)
    of $\beta_t$.}
  \label{fig:sim-plots-medium}
\end{figure}

\begin{figure}[!htb]
  \centering
  \includegraphics[width=1.0\linewidth]{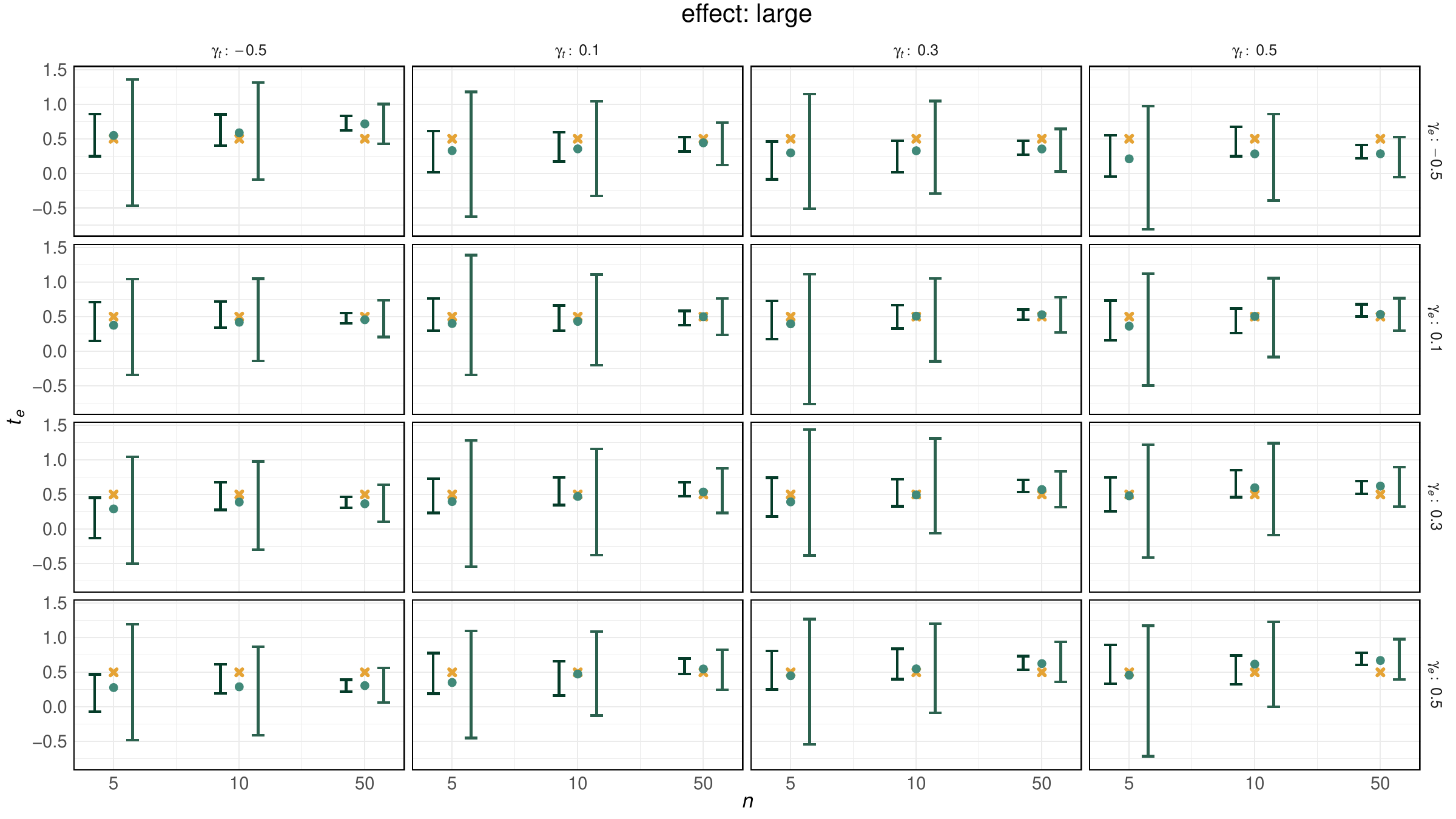}
  \caption{A summary of the results of running \autoref{algo:sim} for some parameter combinations in \autoref{tab:sim-params} and \emph{large} ground truth effect $t_e = 0.5$. The plots report, for each sample size 5, 10, 50,
    the ground truth $t_e$ \gtmark,
    the mean estimate \estmark,
    the 50\% probability interval \errorbar[degascolg] (shifted left),
    and the 95\% probability interval \errorbar[degascolf][0.9ex] (shifted right)
    of $\beta_t$.}
  \label{fig:sim-plots-large}
\end{figure}

\begin{algorithm}[!tb]
\KwIn{parameters $b_s, b_s, k_e, k_s, o_e, e_t, s_e, s_t, t_e, \gamma_e, \gamma_t, n, n_{\textrm{sim}}$}
\KwOut{estimate $\beta_t$; 50\% interval $\ell_{50} \ldots u_{50}$; 95\% interval $\ell_{95} \ldots u_{95}$}
$\mathit{est} \gets \emptyset$\\
\tcp{repeat $n_{\textrm{sim}}$ times}
\For{$r \gets 1 \ldots n_{\textrm{sim}}$}{
  $\mathit{sim} \gets \emptyset$\\
  \tcp{collect $n$ random samples from generative model \eqref{eq:gen-model}}
  \For{$s \gets 1 \ldots n$}{
    $B, K, O \gets$ samples from $\dist{Binomial}(1, 0.5)$\\
    $Z \gets$ sample from $\dist{Normal}(0, 1)$\\
    $S \gets$ sample from $\dist{Normal}(b_s B + k_s K, 1)$\\
    $T \gets$ sample from $\dist{Normal}(o_t O + s_t S + \gamma_t Z, 1)$\\
    $E \gets$ sample from $\dist{Normal}(b_e B + k_e K + o_e O + s_e S + t_e T + \gamma_e Z, 1)$\\
  \tcp{$\mathit{sim}[s, V]$ stores the value of variable $V$ in the $s$th sample}
    $\mathit{sim}[s, B], \mathit{sim}[s, K], \mathit{sim}[s, O], \mathit{sim}[s, Z], \mathit{sim}[s, S], \mathit{sim}[s, T], \mathit{sim}[s, E] \gets B, K, O, Z, S, T, E$
  }
  $f \gets$ fit(\eqref{eq:inf-sim}, $\mathit{sim}$)
  \tcp{fit model \eqref{eq:inf-sim} with data $\mathit{sim}$}
  \tcp{$\mathit{est}[r]$ stores the fitted model's estimate of $\beta_t$ in the $r$th repetition}
  $\mathit{est}[r] \gets $ estimate($\beta_t, f$)
}
\tcp{compute statistics over all $n_{\textrm{sim}}$ repetitions}
$\beta_t \gets$ mean$(\mathit{est})$ \tcp{average}
$\ell_{50}, u_{50} \gets$ HPDI$(\mathit{est}, 0.50)$ \tcp{50\% probability interval} 
$\ell_{95}, u_{95} \gets$ HPDI$(\mathit{est}, 0.95)$ \tcp{95\% probability interval}
\Return{$\beta_t, \ell_{50}, u_{50}, \ell_{95}, u_{95}$}

  \vspace{7pt}
  \caption{Simulation algorithm to assess the sensitivity of the estimate of $T \to E$'s effect on the confounding effect of $Z$ in \autoref{fig:UDAG-adjusted}.}
  \label{algo:sim}
\end{algorithm}

\subsubsection{Simulation Results}

Figures~\ref{fig:sim-plots-small}, \ref{fig:sim-plots-medium}, and \ref{fig:sim-plots-large} display (a significant subset of)
the simulation results.
Each figure includes 16 $= 4 \times 4$ subplots for each combination
of the following values for $\gamma_e$ and $\gamma_t$: -0.5, 0.1, 0.3, 0.5.
\autoref{fig:sim-plots-small} refers to the experiments where the effect $t_e$
to be estimated is small (0.1);
\autoref{fig:sim-plots-medium} where $t_e$ is medium (0.3);
and \autoref{fig:sim-plots-medium} where $t_e$ is large (0.5).
Overall, the subset of experimental results displayed in the three figures
is sufficient to see the main trends
in the sensitivity analysis; anyway,
the replication package~\cite{omitted-vars-replication}
includes plots for all parameter combinations.
The capsule summary of these results
is that it is possible to retrieve
a reasonably precise estimate of the effect $t_e$
provided the confounding effect of the unmeasured $Z$
(parameters $\gamma_e$ and $\gamma_t$)
is not large compared to $t_e$.

Let's first look at the plot grid in \autoref{fig:sim-plots-medium},
which correspond to a ``ground truth'' medium effect $t_e = 0.3$ (marks \gtmark).
When the confounding effects are small ($\gamma_e, \gamma_t = 0.1$, second row or column of the grid)
or medium ($\gamma_e, \gamma_t = 0.3$, third row or column of the grid),
the estimates (marks \estmark) are quite close to the actual effect,
nearly overlapping it.
Remarkably, this holds even for only 5 or 10 samples;
however, a small sample size leads to very wide probability intervals,
even for 50\% probabilities (left of the estimate).
Increasing the sample size to 50 substantially shrinks
the probability intervals;
if obtaining a substantial number of datapoints
is challenging in practice,
these results indicates that
a substantial uncertainty about the accuracy of the estimate would remain.
In contrast,
as we consider larger confounding effects ($\gamma_e, \gamma_t = \pm 0.5$, first and last row or column of the grid),
there is a substantial gap between
the estimates \estmark and the actual effect \gtmark.
Precisely, when $\gamma_e$ and $\gamma_t$ have the same sign
(both positive, or both negative),
$Z$'s bias results in \emph{overestimating} the ground truth effect;
conversely, when $\gamma_e$ and $\gamma_t$ have opposite sign,
the bias results in \emph{underestimating} the ground truth effect.

If we now consider
a ``ground truth'' small effect $t_e = 0.1$
(plot grids in \autoref{fig:sim-plots-small})
or a large effect $t_e = 0.5$
(plot grids in \autoref{fig:sim-plots-large})
we largely see the same trends.
On the one hand, a large effect size
is not strictly ``harder to bias'';
that is, the estimate of $t_e = 0.5$
is biased in a roughly similar way
by certain confounding effects $\gamma_e, \gamma_t$
as the estimate of $t_e = 0.1$.
Intuitively,
this happens because we still condition on all other variables
in the adjustment set, and hence what is left is
the net confounding effect of omitting $Z$ over the estimate.
On the other hand,
if the ``true'' effect we are estimating
is small compared to the biasing influence of $Z$,
the same absolute amount of bias translates
into an estimate error that is possibly much more consequential:
even the 50\% probability intervals clearly overlap zero;
thus, in a real setting
it would be hard to conclude that a definite (positive or negative)
effect exists at all.

\subsection{Sensitivity Analysis with E-values}
\label{sec:tipping-evals}

The sensitivity analysis based on simulation
presented in \autoref{sec:sensitivity-E-T}
is informative and flexible,
but it also has clear disadvantages:
it can be very time consuming,
and it does usually requires a good amount of
empirical data to tune the simulation parameters
in a way that is representative of the domain.
As a less demanding alternative, this section
presents a tipping-point analysis similar to the
one we described in \autoref{sec:lang:quantifying}
for the other illustrative example.

\subsubsection{E-values}

\autoref{sec:lang:quantifying}'s sensitivity analysis
uses the \SMD (scaled-mean difference) as a standardized measure
of the effect of an unmeasured confounder on the treatment.
The \SMD \eqref{eq:SMD} is defined based on
a dichotomous partition of the treatment variable.
In \autoref{sec:intro-confounding}'s domain,
the treatment was the programming language, which is naturally dichotomous.
In contrast, team size $T$ (the treatment variable in our current domain)
is an intrinsically quantitative (numeric) variable;
in order to calculate an \SMD of $Z \to T$,
we would have to arbitrarily partition teams into small vs.\ large.
Instead, we rely on a different kind of tipping-point analysis
based on the notion of E-value~\citep{Evalue2},
which is also applicable to continuous quantitative exposure variables.\footnote{
  We rely on the R package \texttt{EValue}~\citep{EValue-package},
  which implements a variety of
  state-of-the-art sensitivity analysis techniques for unmeasured confounding
  based on the notion of E-value~\citep{Evalue1,Evalue2,Evalue3}.
}

The E-value\footnote{The ``E'' stands for ``Evidence''.}
is ``the minimum strength of association, on the risk ratio scale, that an unmeasured confounder would need to have with both the treatment and the outcome to fully explain away a specific treatment-outcome association, conditional on the measured covariates''~\cite{Evalue2}.
In our scenario,
an estimate $\beta_t$ of the observed effect $T \to E$,
obtained by fitting model \eqref{eq:inf-sim} on empirical data,
would represent the
``specific treatment-outcome association, conditional on the measured covariates.''
The E-value can be computed from $\beta_t$,
as well as $\epsilon_t$ (the standard error of $\beta_t$'s estimate),
an estimate of $\sigma$ (parameter $\sigma$ in model \eqref{eq:inf-sim}),
and a parameter $\delta$ that quantifies the arbitrarily chosen
change in treatment variable $T$.
With these parameters, an E-value of $e$
can be interpreted as follows:
if the unmeasured confounder $Z$
were strong enough to increase, by a factor of $e$,
the probability of raising the exposure $T$ by $\delta$---while simultaneously affecting the outcome $E$
by a similar amount in standardized units---then the observed effect $\beta_t$ would be entirely
due to $Z$'s confounding.

Values $\beta_t$, $\epsilon_t$, and $\sigma$
all come from fitting model \eqref{eq:inf-sim} on empirical data;
this makes the E-value a convenient way of estimating the effect of unmeasured confounders,
since it is a byproduct of a standard regression analysis.
Since it also depends on $\epsilon_t$ and $\sigma$,
computing the E-value from a regression analysis
brings the additional advantage that it takes into account
the uncertainty in the estimates, as well as the other covariates
the regression model conditions on.
In contrast, parameter $\delta$ in the computation of the E-value
can be chosen arbitrarily so that it reflects a variation ``of interest'' in $T$.
Ultimately, $\delta$ still implicitly introduces a dichotomous partition
of the treatment, but it does so in a way that is more apt for a continuous treatment.

\subsubsection{Computing E-values}

Let's get into computing the E-value in our illustrative example of team size and effort.
Since we don't have real-world data directly about the
variables of \autoref{fig:UDAG-adjusted},
we resort, once again,
to simulation to get some plausible data in a convenient format.
Unlike in \autoref{sec:sensitivity-E-T},
now we don't need to perform many repetitions with small sample size,
since the simulated data will not be used for a sensitivity analysis
but to create a synthetic dataset
based on the \autoref{tab:sim-params}'s parameters.
Thus,
we simply draw ten thousand datapoints by sampling model \eqref{eq:gen-model}
for each of the following parameter combinations:
\begin{itemize}
\item $b_e, b_s, k_e, k_s, o_e, o_t, s_e, s_t, t_e$
  are as in \autoref{tab:sim-params}; namely, we consider a range of positive
  effect sizes for $t_e$, whereas we stick with realistic values for the other parameters.

\item $\gamma_e = \gamma_t = 0$;
  in other words, we do not introduce any confounding in the simulated model.
  This simplifying assumption does not affect the following
  analysis,
  since it still makes sense to compute an E-value in such a scenario:
  the E-value quantifies the magnitude of a hypothetical confounder given an observed effect;
  it is not a way of detecting confounders but of reasoning about their possible strength.\footnote{For completeness sake, the replication package includes a computation of E-values in scenarios where $\gamma_e$ and $\gamma_t$ are non-zero and range over the same values as in \autoref{tab:sim-params}.}
\end{itemize}

\begin{figure}[!tb]
  \centering
  \begin{subfigure}{0.72\linewidth}
    \includegraphics[width=\linewidth]{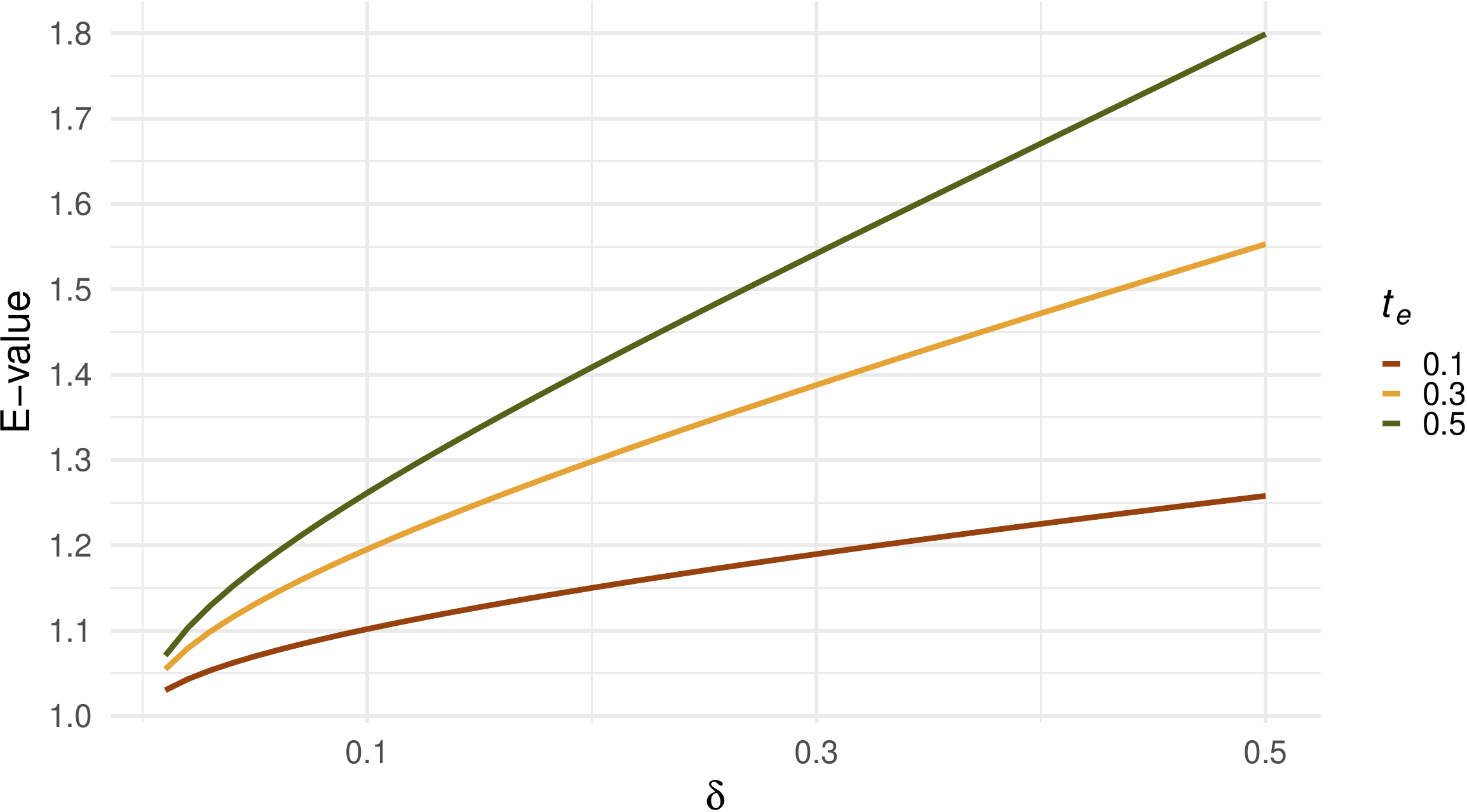}
    \caption{Each colored line shows the E-value for each $\delta$ when
      the $T \to E$ effect is $t_e$.}
    \label{fig:evalue-plot}
  \end{subfigure}
  \hfill
  \begin{subfigure}{0.25\linewidth}
    \begin{tabular}{r rrr}
      \toprule
      & \multicolumn{3}{c}{$\delta$} \\
      \cmidrule(lr){2-4}
      \multicolumn{1}{c}{$t_e$}
      & \multicolumn{1}{c}{$\textsl{0.1}$}
      & \multicolumn{1}{c}{$\textsl{0.3}$}
      & \multicolumn{1}{c}{$\textsl{0.5}$}  \\
      \cmidrule(lr){1-1}
      \cmidrule(lr){2-4}
\multicolumn{1}{r|}{$\textsl{0.1}$} & $1.10$ & $1.19$ & $1.26$ \\
      \multicolumn{1}{r|}{$\textsl{0.3}$} & $1.20$ & $1.39$ & $1.55$ \\
      \multicolumn{1}{r|}{$\textsl{0.5}$} & $1.26$ & $1.54$ & $1.80$ \\
      \bottomrule
    \end{tabular}
\caption{E-values for different combinations of $T \to E$ effect $t_e$ and parameter $\delta$.}
    \label{tab:evalues}
  \end{subfigure}
  \caption{E-value tipping-point analysis for the unknown confounder $Z$ on the $T \to E$
    relation in \autoref{fig:UDAG-adjusted}.}
  \label{fig:evalues-TE}
\end{figure}

Then, we fit model \eqref{eq:inf-sim}
with each of the three simulated datasets $D^{0.1}, D^{0.3}, D^{0.5}$
(one for each value of $t_e = 0.1, 0.3, 0.5$).
The fit with data $D^s$ gives values $\beta_t^{s}$, $\epsilon_t^{s}$, $\sigma^{s}$,
respectively 
of the estimated effect $T \to E$,
of the standard error of this estimate,
and of the standard deviation of model \eqref{eq:inf-sim}'s likelihood.
With $\beta_t^{s}$, $\epsilon_t^{s}$, $\sigma^{s}$,
we compute the E-value for many different values of parameter $\delta$ ranging from 0.01 to 0.5.

\subsubsection{Confounding Sensitivity}

\autoref{fig:evalue-plot} plots the E-values for the
three effect sizes $0.1$, $0.3$, and $0.5$.
Overall, the E-value is proportional to $\delta$
(the parameter that captures the increase in treatment level caused by a possible confounder $Z$)
and to $t_e$ (the actual effect $T \to E$).
This simply reflects that the bigger the effect to be explained away
by a confounder, the stronger the confounder has to sway treatment and outcome
at the same time.

Then, \autoref{tab:evalues}
lists precise E-values for certain combinations of $\delta$ and $t_e$.
Let's look into a couple of interesting parameter combinations:
\begin{enumerate}
\item \label{p:eval-SS}
  When $t_e = \delta = 0.1$,
  the E-value is $1.10$, which can be interpreted as follows:
  if $Z$ is such that, in nominal conditions,
  it can raise by at least 10\%
  the probability of increasing a project's team size by $0.1$---while correspondingly increasing the effort by a ``similar'' amount---(where the increase is subject to some standard distributional assumptions),
  then $Z$ would be sufficient to explain entirely the observed effect $t_e = 0.1$.
  In \eqref{eq:gen-model}'s standardized model,
  $0.1$ is a \emph{small} increase;
  thus, it is not unrealistic that a certain unmeasured factor increases (by 10\%)
  the probability of ending up with a moderately larger team size and effort.

\item \label{p:eval-SL}
  When $t_e = 0.5$ and $\delta = 0.1$,
  the E-value is $1.26$.
  Informally, to explain away a \emph{large} observed effect,
  $Z$ should increase by 26\% the
  probability of introducing a \emph{small} increase in team size ($\delta = 0.1$).
  This is definitely a more significant impact than in the previous point \autoref{p:eval-SS},
  but it is perhaps still plausible.

\item \label{p:eval-LL}
  When $t_e = \delta = 0.5$,
  the E-value is $1.80$.
  Informally, to explain away a \emph{large} observed effect,
  $Z$ should increase by 80\% the
  probability of introducing a \emph{large} increase in team size.
  This scenario is no longer so plausible:
  it is a case of ``extraordinary claims
  [requiring] extraordinary evidence''~\cite{sagan1986broca}.
\end{enumerate}
Comparing scenarios \autoref{p:eval-SL} and \autoref{p:eval-LL}
is also interesting.
Overall, they suggest that it is still possible
that a confounder is responsible for a large effect;
however, this is plausible only if
the confounder $Z$ can bias the estimate of the $T \to E$
effect with only a small change in the treatment $T$ (scenario \autoref{p:eval-SL},
where the $T \to E$ is more sensitive),
as opposed to a large change
(which is less likely to have been missed by the quantitative analysis).

As usual, with a better understanding of the domain
(for example, the typical project characteristics in the company
where the analyzed data was collected),
the E-value analysis could produce more actionable results.
For example,
if one is confident that the estimates of team size
and project effort that are normally produced are usually precise,
it would indicate that it's unlikely that a confounder
would go undetected even for small changes of treatment/outcome
(e.g., $\delta = 0.1)$.
Conversely, if the characteristics of the projects that are
being analyzed make estimates intrinsically imprecise or uncertain,
an unmeasured confounder that substantially increases the probability
of swaying such estimates is a plausible possibility.
As one example, \citet{morasca01prod} study of productivity
(one of those surveyed in \citet{feldtSFT23sysrev})
presents estimates of productivity with a large dispersion
($\sigma \in [0.35, 7.2]$);
these indicate much uncertainty,
thus raising the possibility of unmeasured confounders.

\section{Dealing with Omitted Variable Bias}
\label{sec:discussion-summary}
\label{sec:disc}

This section is a high-level summary of the techniques presented in the paper.
The summary also serves as a procedural checklist,
presenting the main steps of an analysis of confounding
and the order in which they should be followed.

\paragraph{Variables.}
The first step is surveying the variables that characterize the target of our study,
their types (categorical, ordinal, numeric, \ldots),
how costly they are to measure
(e.g., they can be mined from software repositories
vs.\ they require running a controlled experiment),
and how much uncertainty we expect in their measures.
We should also select which variables are the \emph{treatment} (a.k.a.\ \emph{exposure})
and the \emph{outcome}, whose relation is the main focus of the study.
This step underlies all the following ones,
as it provides a way of becoming familiar with the study's domain in an incremental fashion.

\paragraph{Causal DAG.}
The variables of interest, identified in the previous step, serve
as nodes of a DAG such as those
in \autoref{fig:language-quality-skill} and \autoref{fig:productivity-dag}.
Causal DAGs are the notation of choice to succinctly express structural, causal
relations among variables.
As we have demonstrated in the paper,
there are several analyses that are based on a DAG's structure.

What kind of information we can use to build a DAG depends on the
domain we are investigating, and on the maturity of the state of the art.
If there are plenty of rigorous primary studies about the quantities of interest,
and perhaps even systematic reviews or meta-studies,
we can summarize their evidence in a DAG---similarly to what is proposed
by \citet{feldtSFT23sysrev}.
If our study's target is novel or less established,
we may have to rely on domain expertise and intuition to build a DAG.
Even if there is a lot of uncertainty about the precise causal structure
underlying a certain domain,
there are techniques to (partially) validate candidate DAGs~\citep{furiaTF23causal}.
It is quite natural to also consider different possible DAGs,
and to use them to perform a ``what if'' analysis in different scenarios.
In this step of the analysis,
causal DAGs are mainly a convenient notation
to rigorously express our knowledge or hypotheses about the causal relations
among variables,
and to investigate their consequences on the overall results of our analysis.

\paragraph{Adjustment sets.}
As shown in \autoref{sec:adjustment-sets},
given a DAG and treatment\slash outcome variables,
one can systematically compute an \emph{adjustment set}:
a set of covariates that should be conditioned on in a regression
to ensure that the coefficient associated with the treatment variable
estimates the unbiased effect of the treatment on the outcome
(``controlling for'' the spurious influence of any confounders).

In the best-case scenarios,
one can simply use an adjustment set to prevent confounding.
Unfortunately, this is not always possible.
First, the adjustment set's validity is predicated on the accuracy and completeness
of the DAG: if we missed some relevant variables, or misrepresented
some causal relations, an adjustment set no longer guarantees an unbiased estimate.
Second, even if we are confident the DAG is accurate,
certain DAGs do not admit adjustment sets
(because different kinds of confounding
require incompatible adjustment sets~\citep{mcelreath2020statistical}),
or some variables in the adjustment set
are hard or impossible to measure
(a scenario that we explored in \autoref{sec:unmeasured-confounders}).
In these cases, the next steps in this list
can help deal with these shortcomings.

\paragraph{Ballpark estimate of parameters.}
In order to proceed with a sensitivity analysis of
unmeasured confounders,
one needs to collect a rough estimate of the strength
of the main relations among variables in the DAG.
If a good amount of data is available from the study's domain,
we can use them to come up with estimates on the natural scale.
Otherwise, we can still resort to mocking an ersatz model,
based on the DAG's relations, that uses standardized variables.
On a standardized scale, it is easier to make ``guesstimates''
about plausible parameter values (e.g., small vs.\ large),
and to explore the impact of different parameter
combinations---as we did to select the parameter values
in \autoref{tab:sim-params}.

\paragraph{Sensitivity analysis.}
Using the parameter estimates identified in the previous step,
one can perform different kinds of sensitivity analyses of unmeasured confounder.
These analyses provide a quantitative estimate (usually on a normalized scale)
of how much some unmeasured confounder may bias the estimate of an effect of interest.
In the paper, we demonstrated two kinds of so-called \emph{tipping-point} sensitivity analysis,
which express how strong an unmeasured confounder should be to
cancel out an observed treatment\slash outcome effect.
\autoref{sec:lang:quantifying} presented a tipping-point analysis based
on an estimate of the scaled-mean difference of an unmeasured confounder on the treatment,
which is applicable to dichotomous treatment variables.
\autoref{sec:tipping-evals} presented a sensitivity analysis based on
E-values---a probability ratio between the confounded and non-confounded scenarios---which is also applicable to continuous treatment variables.

If a sensitivity analysis indicates that
possible unmeasured confounders are unlikely to exist,
or to have a noticeable impact,
one can proceed with the real data analysis,
reassured that confounding is a remote possibility.

\paragraph{Simulation analysis.}
If the previous step's sensitivity analysis
is inconclusive, in that it failed to rule out
the possibility of confounding,
one can perform a more precise analysis of confounding 
based on simulation.
\autoref{sec:sensitivity-E-T}
illustrated this on our second illustrative example of teamwork productivity.
A simulation analysis is flexible,
because one can explore many different variants of generative
and inference models.
It also supports analyzing the impact of dealing with
small sample sizes in a way that realistically reflects the availability of data
in the study domain.
These advantages come with a cost in terms of simulation time;
usually, however, it still is much cheaper to perform a detailed simulation
than to embark ``blind'' in running an empirical study
without a clear understanding of the possible confounders,
and of the threats to validity they may introduce.

\paragraph{Study design and execution.}
All previous steps are ultimately a preparation for
the design and actual execution of the envisioned study.
Precisely, there are three main outcomes of the previous steps:
\begin{description}
\item[\emph{All clear:}] the analysis indicates that confounding is
  not possible (because of the DAG structure), can be prevented
  (using a suitable adjustment set), or is unlikely to have a sizeable impact
  (as shown by the sensitivity analysis). This is the best-case scenario,
  which bodes well for the validity of our study.

\item[\emph{Proceed with caution:}]
  the analysis indicates that confounding is a possibility,
  but, depending on the effects that are in place
  and on the sample size that we may be able to collect, may or may not be consequential.
  In this case, we may still decide to go ahead with our study or,
  more cautiously, we may perform additional preliminary analyses
  (for example using detailed simulations)
  to gauge more precisely the quantitative relations that animate our domain.

\item[\emph{No go:}]
  the analysis indicates that major confounding is unavoidable,
  and that our estimates of effects are likely to be ridden with uncertainty.
  In this case, it may not be worth to proceed with the study as originally intended.
  Instead, we may refocus our goals, and redefine our research questions,
  so as to move them to a scope that is more likely to be productive.
\end{description}

\subsection{Reporting a Sensitivity Analysis}

To encourage researchers to perform sensitivity analyses as part of their quantitative studies,
we outline a few simple guidelines on how to \emph{report} the results of a sensitivity analysis in a paper.
\begin{description}
\item[\emph{DAG:}] The DAG is a succinct summary of the key assumptions that underlie a sensitivity analysis. Therefore, it's
  always recommended to include the DAG in the paper, preferably with a brief justification for its structure---or a reference
  to another publication that justifies it.
\item[\emph{Parameter estimates:}] The other key ingredient for a sensitivity analysis is an estimate of the main relations among
  variables---in particular, possible unmeasured confounders.
  Since there is usually a degree of uncertainty about the ``real'' parameter values, one should normally indicate
  a \emph{range} of values for each parameter, rather than a single point estimate. Each estimate should also be accompanied
  by a succinct justification: a study that suggests that value, or, at least, a rationale for an educated guess. \autoref{tab:sim-params}
  shows an example of how the various parameter estimates could be reported in tabular form.
\item[\emph{Tipping plot:}] The outcome of a sensitivity analysis can be visually summarized with a plot of the tipping points
  similar to \autoref{fig:tipping-lang-quality} and \autoref{fig:evalue-plot}. As we argued throughout the paper, a sensitivity
  analysis explores and compares different scenarios. Therefore, its outcome is generally not dichotomous but rather a matter of degree.
  A plot provides a concise, visual summary of ``what if'' scenarios such as:
  ``if the unmeasured confounders are within certain ranges, then their confounding effect would/would not be able to nullify the observed effect''.
\end{description}

\section{Conclusions}
\label{sec:concl}

This paper introduced to the empirical software engineering community
techniques to assess and mitigate the so-called omitted variable bias.
These techniques are
crucially based on a causal model of the relations among variables of interest,
formalized by means the causal DAG notation~\citep{pearl09a@reason}.

First, if the causal structure among variables admits an \emph{adjustment set}
that only includes measurable variables, one can correct for
an omitted variable bias by simply conditioning on all variables in the adjustment set.
If this is not possible,
one can perform a \emph{sensitivity analysis},
whose goal is investigating the impact of unknown confounders.
The paper presented different kinds of sensitivity analyses,
including both so-called \emph{tipping-point} analyses based on
canonical distributional assumptions,
and more precise, but also computationally expensive, analyses based on \emph{simulation}.
We demonstrated these techniques on two illustrative examples---the relation between programming languages and code quality,
and the effect on team size on software development effort---taken from recent statistical analyses of data in these two domains~\citep{FSE,feldtSFT23sysrev}.

The main high-level takeaway of this work
is to \emph{think before you act}.
The most effective way of designing an empirical study
is to start with an elicitation of the causal model(s)
that underlie the phenomena under study,
followed by a systematic and explicit sensitivity analysis of possible confounders.
This will lead to a clearer understanding
of the limitations of a particular study and, in turn,
to a more effective study design---one that is less likely to incur major threats to validity.


\end{document}